\documentclass[final,5p,times,twocolumn,nopreprintline]{elsarticle}

\usepackage{amssymb}

\journal{%
Journal of Systems Architecture%
}

\usepackage{pifont}
\newcommand{\cmark}{\ding{51}}%

\usepackage{multirow}

\usepackage{xspace}

\usepackage{bbold}

\usepackage{xcolor}
\definecolor{myOrange}{RGB}{249,199,79}

\usepackage{tikz}
\usetikzlibrary{shapes.geometric}

\usepackage{array}
\newcolumntype{L}[1]{>{\raggedright\let\newline\\\arraybackslash\hspace{0pt}}m{#1}}
\newcolumntype{C}[1]{>{\centering\let\newline\\\arraybackslash\hspace{0pt}}m{#1}}
\newcolumntype{R}[1]{>{\raggedleft\let\newline\\\arraybackslash\hspace{0pt}}m{#1}}

\usepackage[linesnumbered,ruled,vlined]{algorithm2e}

\SetCommentSty{mycommfont}

\usepackage{soul}
\setuldepth{soul}

\usepackage{booktabs}
\usepackage{colortbl}
\usepackage{subcaption}
\usepackage{amsmath}
\usepackage[export]{adjustbox}

\newcommand{\tFaaSFlow}{FaaSFlow\xspace}

\begin{document}

\begin{frontmatter}



\title{BeeFlow: \ul{B}ehavior Tr\ul{ee}-based Serverless Work\ul{flow} Modeling and Scheduling\\
for Resource-Constrained Edge Clusters}


\author[a]{Ke Luo}
\author[a]{Tao Ouyang}
\author[a]{Zhi Zhou}
\author[a]{Xu Chen\corref{cor}}
\ead{chenxu35@mail.sysu.edu.cn}
\cortext[cor]{Corresponding Author}

\affiliation[a]{organization={School of Computer Science and Engineering, Sun Yat-sen University}, 
            addressline={No. 132, Waihuan East Road, Higher Education Megacenter}, 
            city={Guangzhou},
            postcode=510006, 
            country={China}}

\begin{abstract}
Serverless computing has gained popularity in edge computing due to its flexible features, including
the pay-per-use pricing model, auto-scaling capabilities, and multi-tenancy support.
Complex Serverless-based applications typically rely on Serverless workflows (also known as
Serverless function orchestration) to express task execution logic, and numerous application- and
system-level optimization techniques have been developed for Serverless workflow scheduling.
However, there has been limited exploration of optimizing Serverless workflow scheduling in
edge computing systems, particularly in high-density, resource-constrained environments
such as system-on-chip clusters and single-board-computer clusters.
%
In this work, we discover that existing Serverless workflow scheduling techniques typically assume
models with limited expressiveness and cause significant resource contention. 
To address these issues, we propose modeling Serverless workflows using behavior trees, a novel and
fundamentally different approach from existing directed-acyclic-graph- and state machine-based
models.
Behavior tree-based modeling allows for easy analysis without compromising workflow expressiveness.
We further present observations derived from the inherent tree structure of behavior trees for
contention-free function collections and
awareness of exact and empirical concurrent function invocations.
Based on these observations, we introduce \emph{BeeFlow}, a \ul{b}ehavior tr\ul{ee}-based Serverless
work\ul{flow} system tailored for resource-constrained edge clusters.
Experimental results demonstrate that \emph{BeeFlow} achieves up to
$3.2\times$ speedup in a high-density, resource-constrained edge testbed and
$2.5\times$ speedup in a high-profile cloud testbed,
compared with the state-of-the-art.
\emph{BeeFlow} also demonstrates superior robustness in scenarios with heavy system workloads.

\end{abstract}




\end{frontmatter}







\section{Introduction}{%
Serverless computing, an emerging trend in cloud computing, promises to conceal tedious and
error-prone system administration operations from developers to simplify cloud programming
\cite{DBLP-arxiv19-berkeley}.
With appealing features like the pay-per-use pricing model and auto-scaling capabilities,
Serverless computing has caught the interest from both industrial
\cite{WEB-Amazon-Lambda, WEB-Google-Functions, WEB-Microsoft-Functions, WEB-Alibaba-FC, WEB-Apache-OpenWhisk, WEB-OpenFaaS}
and academia communities
\cite{DBLP-arxiv19-berkeley, ACM-cs21-sjtu, DBLP-jcloudc21-survey}.
Serverless computing today typically consists of \emph{Function-as-a-Service} (FaaS) and
\emph{Backend-as-a-Service} (BaaS), where FaaS offers on-demand event-driven cloud function services
and BaaS provides additional managed cloud services, such as object store, message queues, and
key-value store databases%
\footnote{Unless otherwise specified, this work indicates the FaaS part by Serverless computing.}
\cite{DBLP-cacm21-survey}.
Furthermore, to express complex task execution logic, Serverless workflows
\cite{WEB-Amazon-ASL} (also known as Serverless function orchestration) are increasingly being adopted
in Serverless-based applications, and there have been numerous application- and system-level
optimization techniques \cite{DBLP-socc20-wukong,DBLP-socc21-atoll,DBLP-asplos22-FaaSFlow}
developed for Serverless workflow scheduling.

Meanwhile, the growing edge computing infrastructure \cite{10163866,10128791}
also opens new cases for Serverless computing
\cite{DBLP:conf/edge/RajputKCWK22,DBLP:conf/eurosys/LyuCAP022},
since capabilities like on-demand resource-efficient processing and multi-tenancy support are
highly-desired.
In particular, a new form of edge server has been proposed recently
\cite{DBLP:conf/ieeesec/XuZW22}, where multiple system-on-chips (SoCs) or
single-board-computers (SBCs) are interconnected as an SoC/SBC-cluster within a compact rack chassis.
Preliminary evaluations \cite{DBLP:journals/corr/abs-2212-12842} indicate that SoC/SBC-clusters
can achieve great energy efficiency, ownership-cost efficiency, and workload throughput in certain
applications like video transcoding. This form of edge servers presents further chances
for Serverless computing, as SoC/SBC-level workload granularity should be achieved for scheduling.
However, efforts dedicated to Serverless workflow scheduling in edge computing systems are
insufficient, particularly for resource-constrained nodes in SoC/SBC-clusters.

In order to study Serverless workflow scheduling in resource-constrained edge clusters, we
first deploy state-of-the-art Serverless workflow scheduling in a high-density ARM64
SBC-cluster. Benchmark results with 8 representative workflows
reveal that the scheduling optimization deteriorates
in the SBC-cluster, as its eager strategy soon introduces overwhelming workloads in a few
resource-constrained nodes, leading to significant resource contention.
Besides, existing optimization techniques for Serverless workflow scheduling mostly assume
direct-acyclic-graphs (DAGs) or DAG variants
\cite{DBLP-ppopp22-Mashup, DBLP-socc21-atoll, DBLP-socc20-wukong, DBLP-socc18-sprocket},
which restrict task execution logic like looping and retrying, largely limiting the expressiveness of
workflows for various edge applications.

To address the expressiveness and resource contention issues, we propose modeling
Serverless workflows using behavior trees \cite{DBLP-ras22-survey},
as an alternative to existing DAG- and state machine-based models
\cite{WEB-Amazon-Step, WEB-Google-workflows, WEB-Microsoft-logic, WEB-CNCF-workflow}.
Behavior trees are inherently trees, which can be regarded as DAGs with special semantics, thus
allowing existing advances in Serverless workflow scheduling to be exploited
without limiting expressiveness.
%
Moreover, we observe that the tree structure is useful for deriving
contention-free function collections and estimating exact and empirical concurrent
function invocations.

Based on these observations, we introduce \emph{BeeFlow}, a \ul{b}ehavior tr\ul{ee}-based
Serverless work\ul{flow} system tailored for resource-constrained edge clusters.
\emph{BeeFlow} adopts a novel resource allocation approach that uses the contention-free function
collections as resource control units. Additionally, it mitigates intra-workflow resource contention
and balances both intra- and inter-workflow resource contention across nodes through contention-aware
behavior tree partitioning and placement.

\textbf{Contributions.} The main contributions of this work can be summarized as follows:
\begin{enumerate}
    \item[(1)] We propose modeling Serverless workflows using behavior trees.
        The key motivations behind using this new modeling approach are two-fold: first, behavior
        trees are inherently trees, enabling easy analysis
        without compromising expressiveness; second, the tree
        structure can potentially unveil valuable insights into intra-workflow resource contention.
        We also demonstrate that behavior trees are compatible with existing workflows modeled using
        DAGs or state machines.
        To the best of our knowledge, this is the first proposal of behavior trees in Serverless
        computing.
    \item[(2)] We propose \emph{BeeFlow}, a \ul{b}ehavior tr\ul{ee}-based Serverless work\ul{flow}
        system designed for resource-constrained edge clusters.
        \emph{BeeFlow} leverages the observations derived from the behavior tree-based modeling
        to enable contention-aware workflow partitioning and placement, which
        significantly reduces resource contention in constrained edge nodes.
        \emph{BeeFlow} also provides a minimalist reference architecture design to
        facilitate easy integration with existing software or orchestration systems in
        edge computing.
    \item[(3)] Our rigorous experiments reveal that \emph{BeeFlow} effectively mitigates and balances
        resource contention across nodes, achieving
        up to $3.2\times$ speedup in a high-density, resource-constrained edge testbed and
        $2.5\times$ in a high-profile cloud testbed, compared with the state-of-the-art.
        \emph{BeeFlow} also demonstrates superior robustness in scenarios with
        heavy system workloads.
\end{enumerate}

The remainder of this paper is organized as follows.
Section~\ref{sec:motivation} describes the background and detailed motivations for this work.
Section~\ref{sec:scheduling} provides a primer on behavior trees, presents novel observations, and
demonstrates the compatibility.
Section~\ref{sec:architecture} introduces \emph{BeeFlow}, including both the algorithm and
architecture designs,
followed by Section~\ref{sec:evaluation}, which details rigorous experiments.
Finally, Section~\ref{sec:related} reviews related works, and Section~\ref{sec:conclusion}
concludes this work.

}
\section{Background and motivations}{%
\label{sec:motivation}

\subsection{Background}
\label{sec:background}

\textbf{Serverless workflow models.}
Workflow is the most common implementation of internal invocations for orchestrating Serverless
functions with specific business logic \cite{ACM-cs21-sjtu}.
Serverless workflow models can be categorized broadly into two types:
DAG-based
\cite{DBLP-ppopp22-Mashup, DBLP-socc21-atoll, DBLP-socc20-wukong, DBLP-socc18-sprocket}
and state machine-based
\cite{WEB-Amazon-Step, WEB-Google-workflows, WEB-Microsoft-logic, WEB-CNCF-workflow}.
There are also another stream of workflow-as-code frameworks that do not assume explicit models
\cite{WEB-cadence, WEB-temporal}, as exemplified by AWS SWF with Lambda tasks \cite{WEB-swf-lambda}.

In this work, we primarily focus on Serverless workflows with explicit models, as they offer
better trade-offs between flexibility and analyzability, and a brief model comparison is presented
in Section~\ref{sec:model-comparison}.
Nevertheless, workflow-as-code frameworks can still act as the underlying execution engines,
enabling higher-level explicit models to share the advances in existing workflow-as-code
frameworks.

\textbf{Serverless workflow scheduling optimization.}
Recent studies \cite{DBLP-atc18-sand, DBLP-cs19-survey, DBLP-atc21-sonic}
have started to exploit the constrained structures enforced by explicit models
for various application- and system-level Serverless workflow scheduling optimization.
Among them, \emph{FaaSFlow} \cite{DBLP-asplos22-FaaSFlow} proposes
a \emph{worker-side workflow schedule pattern}, in which each workflow is partitioned
into sub-DAGs in a strategic way to exploit instant worker-side function invocations and
in-memory caching.
Specifically, \emph{FaaSFlow} employs a critical path-based analysis for grouping functions and
achieves state-of-the-art performance.

\textbf{SoC- or SBC-cluster for edge computing.}
Meanwhile, there arises a new promising form of edge server, namely
SoC-cluster \cite{DBLP:conf/ieeesec/XuZW22} or SBC-cluster, which orchestrates numerous mobile
system-on-chips (SoCs) or single-board-computers (SBCs) in a compact rack chassis.
SoC/SBC-clusters have been shown to achieve great energy-efficiency, ownership-cost efficiency,
and workload throughput compared with traditional edge servers in certain applications like
video transcoding \cite{DBLP:journals/corr/abs-2212-12842}.
The SoC/SBC-level workload scheduling granularity also makes a good case for Serverless computing.

However, it remains unclear how well existing cloud-oriented advances in Serverless computing
would perform in resource-constrained nodes within these new SoC/SBC-clusters.
To this end, we conduct a preliminary study of Serverless workflow scheduling in an SBC-cluster
and the discussion is deferred to Section~\ref{sec:motivation-study}.

\subsection{Serverless workflow model comparison}
\label{sec:model-comparison}

\textbf{DAG.} With the DAG model, Serverless functions are represented as graph nodes, and
dependencies among them are expressed using directed graph edges.
Relationships between Serverless functions in a general DAG can be
classified into four types \cite{DBLP-middleware20-xanadu}, including
{\large\textcircled{\small{1}}} \texttt{1:1} relationship, where one function depends on
another function,
{\large\textcircled{\small{2}}} \texttt{1:m} relationship, where \texttt{m} functions depend
on one function,
{\large\textcircled{\small{3}}} \texttt{m:1} relationship, where one function depends on
\texttt{m} functions, and
{\large\textcircled{\small{4}}} \texttt{m:n} relationship, which combines the features of
\texttt{1:m} and \texttt{m:1} relationships.
The DAG model is well-suited for static workflows, as it can express any deterministic
dependencies between Serverless functions and is easy to analyze, given that all DAGs are
well-defined, free of cycles/loops, and do not introduce non-determinism
\cite{DBLP-ppopp22-Mashup, DBLP-socc21-atoll, DBLP-socc20-wukong, DBLP-socc18-sprocket}.

\textbf{DAG variants.} Although the DAG model offers several advantages, it cannot express dynamic
logic, including conditioning and looping, which are important for Serverless workflow modeling,
particularly for edge applications in enabling real-time information processing and smart decision
making.
For example, in a Serverless workflow for illegal content recognition \cite{DBLP-asplos22-FaaSFlow},
an image requires processing (e.g., applying mosaic filters) only when it is recognized as
containing illegal content.
In this regard, there are variants of DAG that introduce conditional \texttt{1:m} relationship
\cite{DBLP-middleware20-xanadu}
(only one of \texttt{m} downstream Serverless functions will be triggered based on the result)
or use similar notations \cite{DBLP-asplos22-FaaSFlow}
to enable conditioning in workflows.
While DAG variants introduce dynamism to workflows, conventional analysis methods remain applicable
to them with appropriate adaptations \cite{DBLP-middleware20-xanadu,DBLP-asplos22-FaaSFlow}.

\textbf{Directed-cyclic-graph (DCG) and state machine-based (SM-based) approaches.}
To enable the inclusion of looping logic for functionalities such as dynamic iterations
and retry policies in DAGs, it is necessary to relax the acyclic constraint, resulting in
directed-cyclic-graphs (DCGs) \cite{DBLP:journals/tnsm/WangLSB22}.
Systems or methods based on DCGs \cite{dcg/systems,DBLP:journals/jcloudc/KraemerWA21,DBLP:conf/skg/WuSL11}
often lack support for static analysis or assume specific
usage of cycles, relying on the division of DCGs into sub-DAGs for further operations.
However, to ensure correct state and concurrency management, most commercial platforms
\cite{WEB-Amazon-Step, WEB-Google-workflows, WEB-Microsoft-logic}
have adopted approaches based on state machines, which can be regarded as alternatives to DCGs
in the context of Serverless computing. These approaches expand the well-known finite-state-machine (FSM)
formulation by incorporating nesting to support concurrency
(i.e., a state can encompass multiple child-state machines for concurrent execution).
Although the state machine-based approaches enable all of the deterministic relationships, conditioning,
and looping, they make workflow analysis a daunting task since the transitions among states in a
state machine can be arbitrarily complex.

A comparison among existing Serverless workflow models is summarized in
Table~\ref{tab:model-comparison}, in terms of expressing deterministic relationships, conditioning,
looping, and the merit of easy-analysis. The table indicates that none of existing Serverless workflow
models sufficiently address all aspects, highlighting the need for a new modeling approach.

\begin{table}[!ht]
    \centering
    \caption{Serverless workflow model comparison.}
    \label{tab:model-comparison}
    \vspace{-0.5em}\small
    \begin{tabular}{|r|c|c|c|c|>{\columncolor{red!15}}c|}
        \hline
        & \textbf{DAG}
        & \textbf{DAG variants}
        & \textbf{DCG}
        & \textbf{SM-based}
        \\\hline\hline
        \emph{deterministic}
        & \cmark
        & \cmark
        & \cmark
        & \cmark
        \\\hline
        \emph{conditioning}
        &
        & \cmark
        & \cmark
        & \cmark
        \\\hline
        \emph{looping}
        &
        &
        & \cmark
        & \cmark
        \\\hline
        \emph{easy-analysis}
        & \cmark
        & \cmark
        &
        &
        \\\hline
    \end{tabular}
\end{table}

\subsection{Preliminary study of Serverless workflow scheduling in edge environments}
\label{sec:motivation-study}

In this study, we evaluate the state of the art Serverless workflow scheduling mechanism,
\emph{FaaSFlow} \cite{DBLP-asplos22-FaaSFlow}, in edge environments
using an on-premise SBC-cluster.
The benchmark comprises 8 workflows provided by \emph{FaaSFlow}, including
4 popular scientific workflows from Pegasus workflow executions
\cite{DBLP:journals/fgcs/DeelmanVJRCMMCS15, WEB-pegasus}:
    Cycles (Cyc),
    Epigenomics (Epi),
    Genome (Gen), and
    Soykb (Soy),
and 4 representative Serverless workflow applications:
    Video-FFmpeg (Vid),
    Illegal Recognizer (IR),
    File Processing (FP), and
    Word Count (WC).
It should be noted that the 4 scientific workflows simulate computation time through sleeping,
without performing actual computations.
The SBC-cluster employed for the benchmark is Firefly Cluster Server R1 \cite{WEB-firefly},
a high-density SBC rack server containing 11 ARM64 SBCs in a 1U-rack chassis,
as shown in Figure~\ref{fig:firefly}.

\begin{figure}[!ht]
    \centering
    \includegraphics[scale=0.25]{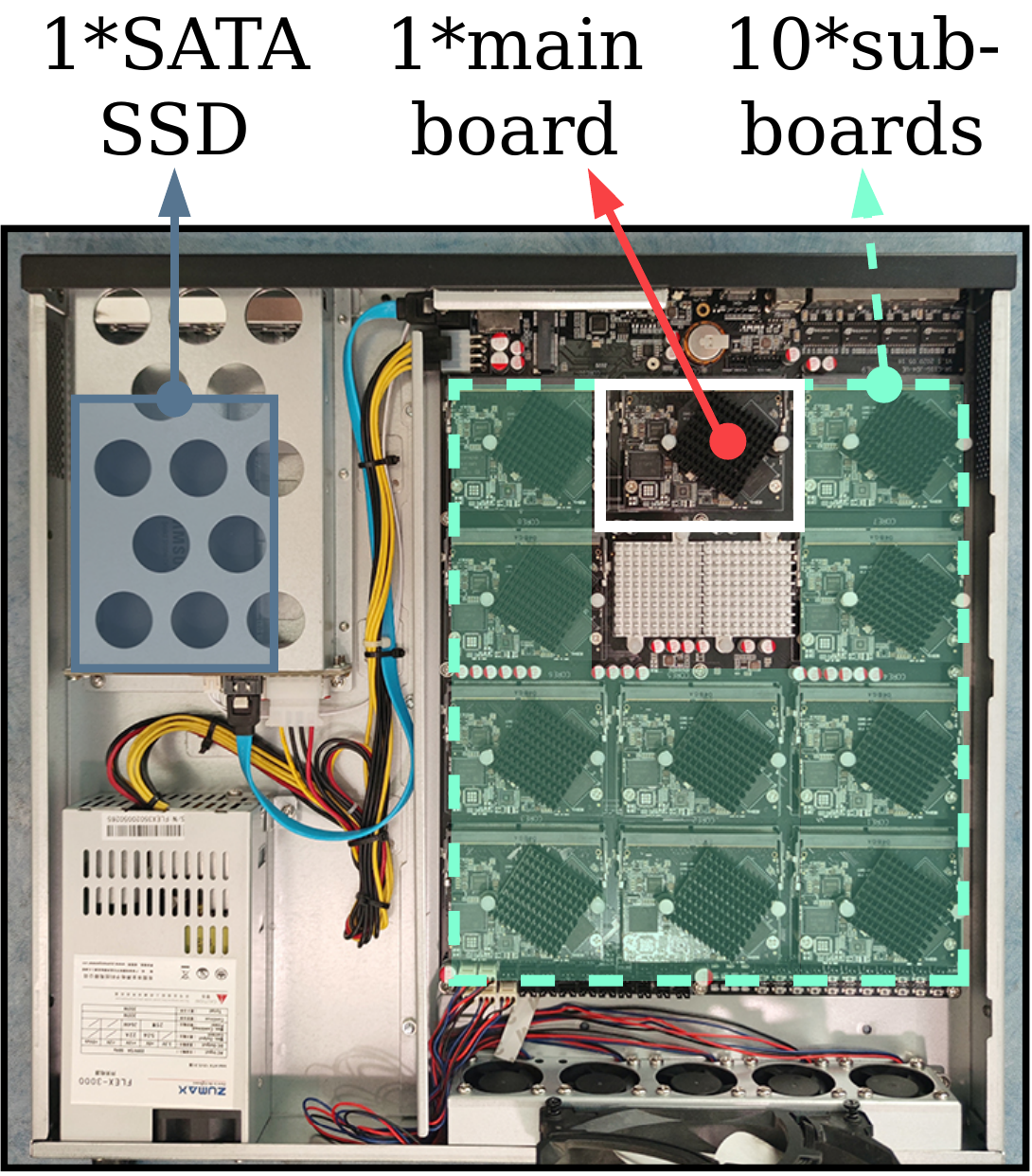}
    \vspace{-0.5em}
    \caption{Bird-view of the FireFly Cluster Server R1.}
    \label{fig:firefly}
\end{figure}

The SBCs within the cluster are interconnected via a 1Gbps switch, and each SBC is equipped with an
RK3399 CPU, 4GB RAM, and 32GB eMMC storage. Additionally, one of the SBCs is designated as the main
SBC and is connected to a 1TB SATA SSD.
For deployment, the main SBC serves as the storage node
(as the additional SSD provides approximately $1.5\times$ better performance than
the internal eMMC storage), and 9 other SBCs serve as worker nodes
(note that 1 SBC is reserved for development).
The detailed specifications are summarized in Table~\ref{tab:firefly}.

\begin{table}[!ht]
    \centering
    \caption{Specifications of the FireFly Cluster Server R1.}
    \label{tab:firefly}
    \vspace{-0.5em}
    \begin{tabular}{@{}R{2.5cm}|L{4.7cm}@{}}
        \toprule
        \textbf{1*main SBC}
        & RK3399 with \mbox{4GB RAM}, \mbox{32GB EMMC}, \mbox{1TB SATA SSD}
        \\\midrule
        \textbf{10*SBC \mbox{\emph{\footnotesize (1 is reserved for dev)}}}
        & RK3399 with \mbox{4GB RAM}, \mbox{32GB EMMC}
        \\\bottomrule
    \end{tabular}
\end{table}

Our benchmark results show that the Cyc workflow fails to pass the tests when deployed in the
SBC-cluster. Upon further investigation,
we identify significant imbalances in data transmission across nodes, indicating that most data
transmission and, consequently, function execution, are handled by a few nodes,
which in turn results in Out-Of-Memory errors.
This disparity in data transmission is also observed in the 7 other successful workflows,
as illustrated in Figure~\ref{fig:distribution}, due to \emph{FaaSFlow}'s strategy of eagerly
co-locating functions that share intermediate results%
\footnote{
    It should be noted that \emph{FaaSFlow} supports adjusting its grouping parameters to behave
    less aggressively. However, this also diminishes the performance gain from its
    \emph{worker-side schedule pattern}, which leads to worse results in our trials.
},
which exacerbates contention in a few nodes.

\begin{figure}[!ht]
    \centering
    \includegraphics[scale=1]{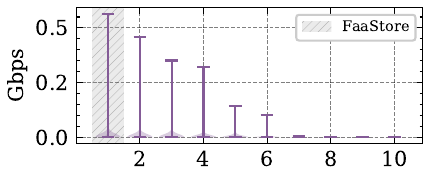}
    \vspace{-0.5em}
    \caption{Per-SBC distribution of 5-second averaged transmission speeds during the tests of
        all workflows except Cyc. \emph{The violin plots are sorted w.r.t. the maximum speed,
        and the SBC serving as the storage node (FaaStore) is marked with a gray shadow.}}
    \label{fig:distribution}
\end{figure}

\subsection{Motivations}

Based on the model comparison and preliminary study above, we identify two major
limitations of existing Serverless workflow modeling and scheduling approaches when applied to
resource-constrained edge computing scenarios:

\textbf{Compromised workflow expressiveness or analyzability.}
There lacks a workflow model that enables easy analysis while supporting both conditioning
and looping logic, which are crucial for accommodating various edge applications.

\textbf{Severe resource contention.}
State-of-the-art Serverless workflow scheduling does not account for the resource-constrained
nature of edge systems such as SoC/SBC-clusters, where resource contention is highly detrimental due to bottlenecks
of all types.

None of the existing works have specifically addressed these limitations. We argue that
a new modeling and scheduling approach is required
to develop optimization techniques without compromising workflow expressiveness and
to alleviate severe contention for resource-constrained edge clusters
simultaneously.

}
\section{Behavior tree-based Serverless workflow modeling}{%
\label{sec:scheduling}
In this section, we propose the use of behavior trees to model Serverless workflows.
We first present an overview of behavior trees,
followed by novel observations drawn from their inherent tree structure.
Finally, we demonstrate the compatibility of behavior trees by establishing connections with
DAG- and state machine-based workflows.

\subsection{A primer on behavior trees}

Behavior tree is a task graph model that organizes tasks into a rooted tree
with a pre-defined set of composites to express task execution logic.
Behavior tree was originally invented and used in computer games to enable modular AI.
Given its hierarchical tree structure that simplifies both synthesis and analysis, behavior tree
has soon become prevalent in robotics control \cite{DBLP-ras22-survey}, inspiring further studies
that employ planning and learning capabilities with behavior trees
\cite{DBLP-icra22-planning, DBLP-aaai21-planning,DBLP-icra21-learning}.

The classic formulation of behavior trees \cite{DBLP-arxiv17-bt} comprises 4 composites:
\emph{sequence}, \emph{fallback}, \emph{parallel}, and \emph{decorator},
as illustrated in Figure~\ref{fig:model:bt}.
These composites are arranged as rooted trees, where each branch
(referred to as a \emph{subtree} hereafter) leads to either a \emph{leaf} or a nested composite.
Specifically, we generalize the \emph{parallel} and \emph{decorator} composites for easier
customization. A detailed description of the \emph{leaf} and composites is provided below:
\begin{enumerate}
    \item[(1)] \emph{leaf}:
        A \emph{leaf} represents a specific task to perform%
        \footnote{
            Classic behavior tree formulations typically categorize \emph{leaves} into two types,
            \emph{action} and \emph{condition}. However, as a \emph{condition} can be
            realized through an \emph{action} that evaluates simple predicates, we do not
            differentiate between them in this work for the sake of clarity.
        }%
        , which may either succeed or fail upon completion%
        \footnote{
            Classic behavior tree formulations also permit an \emph{action} to return a
            \emph{running} state for asynchronous execution. However, since asynchronous execution
            is seldom involved in the context of Serverless workflow modeling, we do not include the
            \emph{running} state in this work. Nevertheless, the observations and algorithms derived
            in this work should be applicable to formulations that include the \emph{running} state
            as well.
        }%
        . In the context of Serverless workflow, a \emph{leaf} may be as simple as invoking a
        Serverless function (we will use \emph{leaf} and \emph{function} interchangeably hereafter).
    \item[(2)] \emph{sequence}:
        A \emph{sequence} composite executes its \emph{subtrees} in order (from left to right in
        the diagrams, by convention). A \emph{sequence} composite succeeds if, and only if, all of
        its \emph{subtrees} succeed. Otherwise, it fails immediately upon any \emph{subtree}
        failure.
    \item[(3)] \emph{fallback}:
        A \emph{fallback} composite executes its \emph{subtrees} in order. Unlike
        a \emph{sequence} composite, a \emph{fallback} composite succeeds and terminates
        immediately once any one of its \emph{subtrees} succeeds. Otherwise, it continues attempting
        the remaining \emph{subtrees} and fails only if none are left.
    \item[(4)] \emph{parallel}:
        A \emph{parallel} composite executes its \emph{subtrees} concurrently. The result of
        a \emph{parallel} composite is determined by a customizable aggregate function
        (\emph{agg}), which can access the results from all \emph{subtrees}.
        A typical aggregate pattern is the ``$m$-out-of-$n$'', where a \emph{parallel} composite
        succeeds if at least $m$ of its $n$ \emph{subtrees} succeed.
        Parallel is the only composite involving concurrency in this formulation.
    \item[(5)] \emph{decorator}:
        A \emph{decorator} composite enhances its unique \emph{subtree}. A \emph{decorator} first
        executes its \emph{subtree}, followed by executing a customizable \emph{tail} function.
        The \emph{tail} function has access to the \emph{subtree} result and decides whether the
        \emph{subtree} should be re-entered or provides a specific result to return.
        By customizing the \emph{tail} function, patterns such as
        ``negating the \emph{subtree} result'',
        ``retrying the \emph{subtree} for at most $n$ times'', and
        ``looping over a list''
        can be easily achieved within the behavior tree model.
        Decorator is the only composite involving repetition in this formulation.
\end{enumerate}

\begin{figure}[!ht]
    \centering
    \includegraphics[scale=1.25]{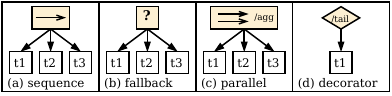}
    \vspace{-0.5em}
    \caption{Behavior tree composites. \emph{``/agg" and ``/tail" are used to indicate customizable functions
        in parallel and decorator composites, respectively. And ``/\{description\}"
        is used to indicate concrete functions implemented for specific purposes in the following figures.}
    }
    \label{fig:model:bt}
\end{figure}

In addition to the above listed features for the \emph{leaf} and composites, a use case of
behavior trees for large language model (LLM)-based domain-specific program generation
\cite{Gupta_2023_CVPR}
is provided in \ref{appendix:real-world} to demonstrate how these components work together
in a real-world scenario.

\subsection{Observations from the inherent tree structure of behavior trees}
\label{sec:observations}

Before discussing our observations, we revise the common path definitions in rooted trees
for behavior trees. First, the raw path function is defined for behavior tree nodes
in a recursive manner:
\begin{equation*}\footnotesize
raw(node) = \begin{cases}
    \{node\}
    ,& node \text{ is root},\\
    raw(node.prev).add(node)
    ,& (node.prev \text{ is not none}) \text{ and}\\
    \hspace{6.3em} (node.parent & \hspace{-1.2em} \text{ is } sequence \text{ or } fallback),\\
    raw(node.parent).add(node)
    ,& \text{otherwise},
\end{cases}
\end{equation*}
where $node.parent$ is the direct parent composite for any non-root node, and
$node.prev$ is not none only if a node has previous sibling nodes
sharing the same direct parent composite.

We further define the path function as
\begin{align*}
path(node) = \left\{
    expand(n) \text{ for } n \text{ in } raw(node)
\right\}.
\end{align*}

The $expand$ utility takes in either a \emph{function} or a composite.
If the input is a \emph{function}, it is directly added into the $path$.
On the other hand, if the input is an ancestor composite of $node$, it is disregarded.
In all other cases, the composite is expanded into a path that leads to any \emph{leaf}
within the context of the \emph{subtree} rooted by the composite. Subsequently, each
associated \emph{function} is added into the $path$.
It should be noted that the path to a node is not necessarily unique,
while the raw path to a node indeed is.

Finally, we define the prefix function as
\begin{equation*}
prefix(node) = \begin{cases}
    raw(node), & node \text{ is composite},\\
    raw(node) \setminus \{node\}, & node \text{ is } function,
\end{cases}
\end{equation*}
where the ``$\setminus$" symbol represents the \texttt{setminus} operation%
\footnote{$\mathcal{A} \setminus \mathcal{B} = \{x \in \mathcal{A}: x \notin \mathcal{B}\}$.},
and state that any order-preserving subset of $path(node)$ is a subpath to the node.

\textbf{Behavior trees enable easy-analysis while retaining full expressiveness.}
As inherently tree structures, behavior trees exhibit hierarchical and modular properties
at all scales, meaning that any branch of a composite is also a regular behavior tree,
which may potentially inspire a new stream of tree-based optimization strategies
for Serverless workflow scheduling.
Although tree structures can be considered as special DAGs, behavior trees convey distinct semantics
from DAGs, thus allowing the expression of conditioning and looping logic using \emph{fallback} and
\emph{decorator} composites, respectively.
Taken together, these aspects indicate that behavior trees enable easy-analysis without limiting
expressiveness.

\textbf{A behavior tree subpath defines an exclusive \emph{function} set.}
By construction, \emph{functions} involved in a subpath are strictly ordered, meaning that
a \emph{function} cannot start before all its precedent \emph{functions} have finished%
\footnote{
    There are cases where \emph{functions} will be skipped due to the presence of \emph{fallback}
    composites. In these instances, we consider ``skipped'' as a special kind of execution.
    Additionally, repetitions generated by a \emph{decorator} composite can be understood as
    a dynamic ``roll-out'' process and won't affect the conclusions.
}.
This full ordering guarantees that all these \emph{functions} are exclusive, implying that at most
one of the \emph{function} can be active at any time. This further suggests that a subpath
can serve as a viable resource control unit \cite{WEB-control-groups, WEB-k8s-namespaces}
to enable contention-free resource sharing among \emph{functions}.

\textbf{The same prefixes indicate identical first-time invocations.}
We assert that two prefixes are the same if they always reduce to the same \emph{function}
sequence in any workflow execution.
In the context of Serverless workflows, the overhead of composites can generally be assumed
negligible compared with \emph{functions}. Consequently, if two \emph{functions} have the same
prefixes, their first-time invocations will be ready simultaneously.
Awareness of exact concurrent invocations is valuable, as transient bursts can negatively impact
systems \cite{USENIX-atc22-rund}, leading to performance degradation due to
optimistic CPU spinning, irregular I/O accesses, network resource contention, and so on.
With this awareness, concurrent invocations can be spread out to mitigate side-effects.

\textbf{Runtime traces can supplement the awareness of exact concurrent invocations.}
Although precise, the concurrent invocations derived from prefixes may be sparse.
To achieve better coverage, we propose instrumenting behavior trees with runtime traces
\cite{WEB-otel}.
Runtime traces track actual workflow executions, allowing for the study of workflow dynamics,
such as
``how long and how likely a \emph{function} will be executed in a workflow request'' and
``how many iterations a \emph{decorator} composite may yield''.
By resolving this dynamicity, empirical concurrent invocations can be estimated even when
\emph{functions} do not share the same prefixes, providing a robust complement to the exact
concurrency.
Furthermore, miscellaneous intra-workflow resource contention can be assessed, depending
on the granularity of the runtime traces.

\subsection{Compatibility of behavior trees}

Although behavior trees consist of entirely different building blocks from conventional models,
we demonstrate how existing DAG- and state machine-based workflows can be converted to behavior
trees, arguing for their sound compatibility.

{Any DAG can be converted to a behavior tree with non-trivial concurrency.}
Figure~\ref{fig:model:cvt} illustrates the overall idea of converting any DAG to a behavior tree.
The conversion follows a recursive method: given a DAG,
{\large\textcircled{\small{1}}} group its source nodes into a \emph{parallel} composite (head) and remove
    them from the DAG,
{\large\textcircled{\small{2}}} group its sink nodes into a \emph{parallel} composite (tail) and remove
    them from the DAG,
{\large\textcircled{\small{3}}} for each connected sub-DAG with the remaining nodes, go through
    steps {\large\textcircled{\small{1}}}-{\large\textcircled{\small{5}}} herein to generate
    a \emph{subtree},
{\large\textcircled{\small{4}}} compose the \emph{subtrees} resulting from step
    {\large\textcircled{\small{3}}} with a \emph{parallel} composite,
{\large\textcircled{\small{5}}} compose the \emph{subtrees} resulting from steps
    {\large\textcircled{\small{1}}\textcircled{\small{4}}\textcircled{\small{2}}}
    with a \emph{sequence} composite.
If any of the conversion steps encounters an empty DAG, that step can be simply omitted.
Additionally, the \emph{agg} function used in the \emph{parallel} composites requires
all \emph{subtrees} to succeed.

\begin{figure}[!ht]
    \centering
    \includegraphics[scale=0.95]{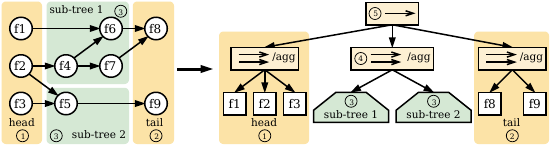}
    \vspace{-0.5em}
    \caption{Overall idea behind DAG-to-behavior tree conversion.}
    \label{fig:model:cvt}
\end{figure}

The conversion method is optimal if DAG nodes are homogeneous, since the conversion steps can be
understood as aligning DAG nodes to the critical path. Even when DAG nodes are heterogeneous, the
conversion method should retain non-trivial concurrency as simultaneously removing the source and
sink nodes tends to divide the remaining nodes into multiple sub-DAGs, which will correspond to
independent \emph{subtrees} that can run in parallel after conversion.

{Any state machine can be converted to a behavior tree with equivalent task execution logic.}
Figure~\ref{fig:model:cvt-sel} demonstrates a selector structure implemented with behavior tree
building blocks, which relies on a \texttt{SEL} variable in the payload to determine the
\emph{subtree}/state to execute and uses another \texttt{END} flag to determine whether the workflow
should terminate. In addition, the initial \texttt{SEL} value is used to decide the start
\emph{subtree}/state.
With the selector structure, any conventional FSMs can be converted to behavior trees with
equivalent task execution logic.
As previously mentioned in Section~\ref{sec:model-comparison}, state machine-based models for Serverless
workflows typically extend FSM by allowing a state to encapsulate multiple child state machines
for concurrent execution. This extension aligns with the logic of the \emph{parallel} composite
in behavior trees, as each branch of a \emph{parallel} composite is also a regular behavior tree
(which in turn can represent a child state machine).
In this regard, we can convert any state machine to a behavior tree using a combination of
the selector structure and \emph{parallel} composites.

\begin{figure}[!ht]
    \centering
    \includegraphics[scale=1.1]{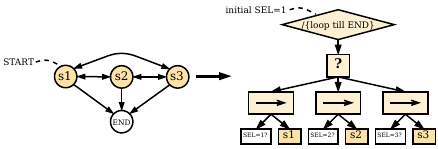}
    \vspace{-0.5em}
    \caption{Overall idea behind state machine-to-behavior tree conversion.
             \emph{``/\{loop till END\}" indicates that the decorator function is implemented
                   to issue re-entrances unless the ``END'' flag is true.}}
    \label{fig:model:cvt-sel}
\end{figure}

However, although automatic conversion enables the generated behavior trees to express equivalent
task execution logic with state machines, the static structure information will be lost during
conversion. This suggests that the automatic conversion should only be used for backward
compatibility, or better conversion methods should be explored. 

}
\section{BeeFlow designs for resource-constrained edge clusters}{%
\label{sec:architecture}

\begin{figure*}[!ht]
    \centering
    \includegraphics[width=0.7\linewidth]{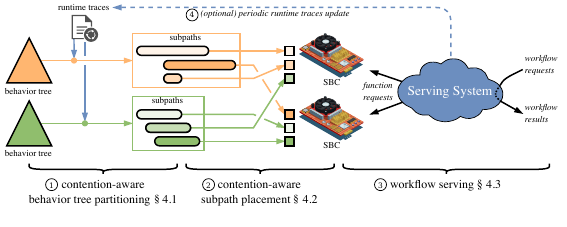}
    \vspace{-0.5em}
    \caption{The high-level pipeline of \emph{BeeFlow}.}
    \label{fig:pipeline}
\end{figure*}

In this section, we introduce \emph{BeeFlow}, a \ul{b}ehavior tr\ul{ee}-based Serverless
work\ul{flow} system specifically designed for resource-constrained edge clusters,
based on the observations derived from the inherent tree structure of behavior trees.
%
Figure~\ref{fig:pipeline} demonstrates the high-level pipeline of \emph{BeeFlow}, which involves
the following steps:
{\large\textcircled{\small{1}}} behavior tree instances are initially partitioned into sets of
    subpaths using a runtime traces-instrumented, contention-aware partitioning algorithm;
{\large\textcircled{\small{2}}} subsequently, subpaths from different behavior tree instances
    are distributed across multiple server nodes via a contention-aware placement algorithm;
{\large\textcircled{\small{3}}} a serving system then translates workflow requests into
    \emph{function} requests, invokes \emph{functions} maintained by subpaths deployed in
    the previous step, and returns the workflow results;
{\large\textcircled{\small{4}}} optionally, runtime traces can be acquired on-the-fly
    and utilized to update the partitioning and placement decisions.

\subsection{Contention-aware behavior tree partitioning}

As observed in Section~\ref{sec:observations}, \emph{functions} involved in a subpath are
exclusive, enabling contention-free sharing of the same set of hardware resources.
To exploit subpath-based scheduling, we first need to partition behavior trees
into subpaths.

\begin{figure}[!ht]
    \centering
    \includegraphics[scale=1]{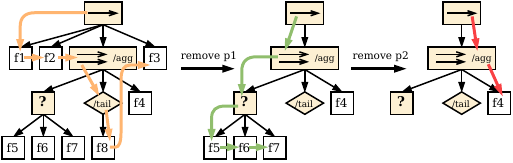}
    \vspace{-0.5em}
    \caption{Example subpath generation for a behavior tree instance.
        \textit{Three subpaths are generated,
            namely sp1 (f1-f2-f8-f3), sp2 (f5-f6-f7), and sp3 (f4).
        }
    }
    \label{fig:beeflow:paths}
\end{figure}

Figure~\ref{fig:beeflow:paths} provides an illustrative example of subpath generation.
In each phase of the example, a subpath is generated, and \emph{functions} involved
in the subpath are removed from the behavior tree for the next phase.
Specifically, in the first phase, subpath \emph{sp1} is generated, involving \emph{functions}
\emph{f1, f2, f8, f3}.
The second phase continues working on the behavior tree resulted from the first phase
and collects \emph{functions} \emph{f5, f6, f7} into subpath \emph{sp2}.
After subpath \emph{sp2} is removed, only \emph{function} \emph{f4} remains and
it is gathered into subpath \emph{sp3}.
After three phases, all \emph{functions} are grouped into separate subpaths.
We assert that the iterative removal process applied in the example is canonical to all
behavior trees, as the ordering among \emph{functions} in behavior trees is transitive,
being invariant to \emph{function} removals.

\begin{algorithm}[!ht]
\caption{I/O-contention-aware partitioning}
\label{alg:splitting}

\newcommand{ \algoDataWorkflow   }{\ul{behavior tree}}
\newcommand{ \algoDataTraces     }{\ul{traces}}
\newcommand{ \algoMathFetchSpeed }{v_{in}}
\newcommand{ \algoMathPutSpeed   }{v_{out}}

\KwData{
    \algoDataWorkflow,
    \algoDataTraces,
    fetch speed $\algoMathFetchSpeed$,
    put speed $\algoMathPutSpeed$
}

\newcommand{ \algoMathResults }{Res}

\KwResult{
    subpaths $\algoMathResults$
}

\newcommand{ \algoVarInitDelay   }{$d_{init}$}
\newcommand{ \algoVarExecDelay   }{$d_{exec}$}
\newcommand{ \algoVarInputSize   }{$s_{in}$}
\newcommand{ \algoVarOutputSize  }{$s_{out}$}
\newcommand{ \algoMathInputSize  }{s_{in}}
\newcommand{ \algoMathOutputSize }{s_{out}}

Estimate
the initialization delay \algoVarInitDelay,
the execution delay \algoVarExecDelay,
the input size \algoVarInputSize, and
the output size \algoVarOutputSize{}
for each function statistically given the \algoDataTraces\;

\newcommand{ \algoVarFetchDelay  }{$d_{in}$}
\newcommand{ \algoMathFetchDelay }{d_{in}}
\newcommand{ \algoVarPutDelay    }{$d_{out}$}
\newcommand{ \algoMathPutDelay   }{d_{out}}

Estimate the fetch delay
$\algoMathFetchDelay \leftarrow \frac{\algoMathInputSize}{\algoMathFetchSpeed}$
and the put delay
$\algoMathPutDelay \leftarrow \frac{\algoMathOutputSize}{\algoMathPutSpeed}$
for each function\;

Estimate the earliest start time for each function given the estimated
\algoVarInitDelay, \algoVarFetchDelay, \algoVarExecDelay, and \algoVarPutDelay,
conforming to the \algoDataWorkflow{}\;

\newcommand{ \algoVarFns    }{$\mathcal{F}$}
\newcommand{ \algoVarPath   }{$p$}
\newcommand{ \algoVarPaths  }{$\mathcal{P}$}
\newcommand{ \algoMathPath  }{p}
\newcommand{ \algoMathPaths }{\mathcal{P}}

\Repeat{the \algoDataWorkflow{} contains no function}{

    Find the timestamp with the maximum I/O contention degree, and get the
    corresponding functions \algoVarFns\;

    \SetKw{Break}{break}
    \While{true}{

        Find all subpaths \algoVarPaths{} 
        involving any function in \algoVarFns{} in the \algoDataWorkflow\;

        \lIf{\algoVarPaths{} is empty}{\Break}

        $\algoMathPath_{max} \leftarrow
        {\arg\max}_{\algoMathPath \in \algoMathPaths}{
            \sum_{\textrm{I/O periods}}(\algoMathPath)
        }$\;

        $\algoMathResults.insert( \algoMathPath_{max} )$\;

        Remove $\algoMathPath_{max}$ from the \algoDataWorkflow\;
    }

}
\end{algorithm}

Moreover, we extend the iterative removal process to promote subpaths with desired
properties. We propose I/O-contention-aware partitioning, since I/O devices
equipped in SBCs in our case are much slower than those in cloud servers and may bottleneck
overall performance.
With I/O-contention-aware partitioning (Algorithm~\ref{alg:splitting}),
subpaths contributing to the maximum I/O-contention degree are pruned first, and
subpaths that yield the most I/O are promoted, bearing in mind that \emph{functions} within
the same subpath are contention-free and more resources can be allocated to subpaths
with higher demands to effectively accelerate the I/O.
More specifically, we assume the lifecycle of a \emph{function} consists of 4 periods:
initialization, input fetching, execution, and output putting, which follows the common idiom for
Serverless functions. When aided with runtime traces, the initialization delay, input size,
execution delay, and output size can be estimated for each \emph{function}, enabling the alignment
of subpaths to examine I/O contention degrees and I/O periods.

Although our proposed partitioning algorithm focuses specifically on I/O-contention, we note that
concerns for promotion can be versatile. For instance, if resource-efficiency is the primary
objective, subpaths with \emph{functions} posing similar resource requests can be
promoted to ensure resources are effectively multiplexed. If order
information is exploited to pre-warm \emph{functions}, subpaths with less
non-determinism may be prioritized. In addition to single-purpose promotion, carefully crafted
polices with mixed purposes can also be effective, suggesting a broad optimization space for
resource-constrained Serverless systems.

\subsection{Contention-aware subpath placement}

Once behavior tree instances are partitioned into subpaths, we can distribute them
across a set of server nodes.
To achieve this, we propose a contention-aware subpath placement strategy, wherein
subpaths contributing more to the intra-workflow contention are scheduled first, and
node assignment is determined so that the overall costs among nodes are balanced while
the delta cost produced by each subpath is minimized. 
The detailed steps are illustrated in Algorithm~\ref{alg:placement}.

\begin{algorithm}[!ht]
\caption{I/O-contention-aware subpath placement}
\label{alg:placement}

\newcommand{ \algoMathChains         }{\mathcal{P}}
\newcommand{ \algoMathWorkflowChains }{\{
    \algoMathChains_1, \ldots, \algoMathChains_m
\}}
\newcommand{ \algoMathNodes          }{\mathcal{N}}
\newcommand{ \algoMathOverlapPenalty }{f_{penalty}}

\KwData{
    $m$ behavior tree instances as subpath sets
        $\algoMathChains_1, \ldots, \algoMathChains_m$,
    server nodes $\algoMathNodes$,
    contention penalty function $\algoMathOverlapPenalty$
}

\newcommand{ \algoMathDecision }{\mathcal{D}}

\KwResult{
    placement decision $\algoMathDecision$
}

\newcommand{ \algoMathWorkflowPlaced }{%
\{\algoMathChains'_1, \ldots, \algoMathChains'_m\}}

Create empty subpath sets $\algoMathWorkflowPlaced$ for each node
in $\algoMathNodes$\;

\newcommand{ \algoMathWeight  }{w}
\newcommand{ \algoMathWeights }{\{w_1, \ldots, w_n\}}

Initialize accumulated costs $\algoMathWeights \leftarrow \{0\}$\ for $\algoMathNodes$\;

\Repeat{$\algoMathChains_1, \ldots, \algoMathChains_m$ are all empty}{

    \newcommand{ \algoMathPath }{p}
    
    Find the subpath that will yield the maximum penalty reduction
    over all behavior trees $\algoMathChains_1, \ldots, \algoMathChains_m$,
    say $\algoMathPath_i \in \algoMathChains_j$ yields the maximum reduction
    $\algoMathOverlapPenalty(\algoMathChains_j)
    - \algoMathOverlapPenalty(\algoMathChains_j \setminus \{\algoMathPath_i\})$%
    \;

    \newcommand{ \algoMathCost }{c}
    
    Calculate the cost of placing $\algoMathPath_i$ into each node in $\algoMathNodes$
    (node index is omitted for clarity),
    which is $\algoMathCost \leftarrow
    \algoMathOverlapPenalty(\algoMathChains'_j \cup \{\algoMathPath_i\})
    - \algoMathOverlapPenalty(\algoMathChains'_j)
    + \infty \cdot \mathbb{1}_{try\_place(\algoMathPath_i) == false}
    $\;

    \uIf{
        $\algoMathWeight_x + \algoMathCost_x > \max\algoMathWeights,
        \forall x \in 1..\vert\algoMathNodes\vert$
    }{
        \tcp{minimize the overall cost}
        sel $\leftarrow \arg\min_{x \in 1..\vert\algoMathNodes\vert}
        \algoMathWeight_x + \algoMathCost_x$\;
    } \Else {
        \tcp{minimize the delta cost}
        sel $\leftarrow \arg\min_{
            x \in 1..\vert\algoMathNodes\vert,
            \algoMathWeight_x + \algoMathCost_x \leq \max\algoMathWeights
        }
        \algoMathCost_x$\;
    }

    $\algoMathDecision[\algoMathPath_i] =$ sel\;
    
    Place subpath $\algoMathPath_i$ into the selected node
    and update the corresponding $\algoMathWorkflowPlaced$ and
    $\algoMathWeight_{sel}$\;

    $\algoMathChains_j = \algoMathChains_j \setminus \{\algoMathPath_i\}$\;
}

\end{algorithm}

In particular, the contention penalty function $f_{penalty}$ leverages the awareness of empirical
concurrent invocations revealed in Section~\ref{sec:observations}, which aligns
all \emph{functions} of the target workflow delegated to the target server node, resulting in intervals
with different I/O contention degrees.
Then, it scales these intervals by the square of the degrees and
finally sums all scaled intervals as the cost of the corresponding workflow.
Additionally, common constraints, such as CPU cores limits and memory size limits, can be enforced
by the \texttt{try\_place} indicator penalty.

\subsection{Reference architecture design}

We propose a reference architecture design accompanying our contention-aware partitioning and
placement algorithms for resource-constrained edge clusters.
The architecture design is intentionally kept simple to facilitate easy integration
with lightweight and portable software or orchestration systems in edge computing \cite{WEB-kubeedge,WEB-k3s}.

First, we adopt the subpath as the deployment unit to enable contention-free resource
sharing among \emph{functions} and introduce a sidecar process to manage corresponding
\emph{functions} for each subpath.
The sidecar subscribes to a message broker to listen for \emph{function} requests.
Upon receiving a request, it triggers the target \emph{function} via shared Inter-Process
Communication (IPC). Once the \emph{function} is completed, the results are published back to
the broker.
Additionally, all \emph{functions} have direct access to a storage service.

Moreover, behavior tree workflows are maintained by gateways, which are responsible
for handling external workflow requests, publishing \emph{function} requests, and
receiving \emph{function} results via the message broker. They also keep track of
the task execution logic of behavior tree workflows.
Importantly, multiple gateways can co-exist to maintain different workflows or
different instances of the same workflow in an independent manner, ensuring that the gateways
do not bottleneck the overall performance.
Figure~\ref{fig:eval:arch} provides an overview of the architecture design.

\begin{figure}[!ht]
    \centering
    \includegraphics[width=0.88\linewidth]{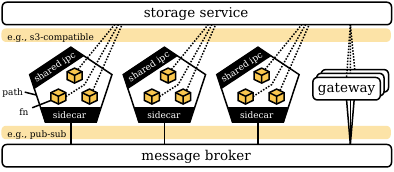}
    \vspace{-0.5em}
    \caption{Reference architecture design for resource-constrained edge clusters.}
    \label{fig:eval:arch}
\end{figure}

\textbf{Function payload and application data.} It is important to note that two services assist in
transferring data between \emph{functions}: the message broker and storage service.
We refer to the data traveling through the message broker as function payload and
the data traveling through the storage service as application data.
The decoupled design for function payload and application data is necessary,
as message brokers typically enforce a strict size limitation (e.g., 1MB
\cite{WEB-NATS-limit, WEB-Kafka-limit})
crucial for achieving low-latency and high-throughput messaging.
However, many applications require the transfer of input, output, or intermediate data
with sizes much larger than the size limitation
\cite{DBLP-socc18-sprocket},
which is why an additional storage service
is introduced.

When writing a piece of data to the storage service, a unique addressable ID is returned.
The writer can then store the ID into the function payload. Any subsequent \emph{functions}
triggered (either directly or indirectly) can access the data via the ID stored in the
payload.
With this two-level design, the distributed deployment of the storage service can be
simplified for resource-constrained edge clusters, where the storage service can directly compose
independent per-worker storage systems for transient, non-critical intermediate data, better
distributing disk and network I/O without incurring extra synchronization costs.

By separating large application data from function payload, data traveling through the
message broker are minimized, ensuring high throughput and low latency. Additionally,
application data are acquired only when necessary, reducing redundant network and disk I/O
to the greatest extent possible.

}
\section{Evaluation}{%
\label{sec:evaluation}
We present the experimental evaluation of \emph{BeeFlow} in two real-world testbeds:
a resource-constrained edge testbed the same as the one we use in the preliminary study
(Section~\ref{sec:motivation-study}), and
a high-profile cloud testbed configured the same as \emph{FaaSFlow}
for fair comparison.
Evaluation results corroborate that \emph{BeeFlow} can outperform
\emph{FaaSFlow} in both edge and cloud settings, particularly with
a higher performance gain in the resource-constrained edge environment.

\subsection{Benchmarking}
\label{sec:benchmarking}

We utilize the 8 workflows provided by \emph{FaaSFlow} for benchmarking, including 4 scientific
workflows and 4 application workflows, as introduced in Section~\ref{sec:motivation-study}.
End-to-end latency tests are conducted for workflows in two modes:
single (sg) and co-run (co).
In the single mode, tests against different workflows are performed without interleaving,
ensuring that different workflows do not interfere with each other.
On the contrary, tests against different workflows are performed simultaneously in the co-run
mode, resulting in more intense mutual resource contention.
For each workflow in both modes, test requests are emitted by a dedicated closed-loop client thread,
i.e., test requests are sent to a workflow one after another.

Tests in the single mode are useful for studying intra-workflow resource contention,
while tests in the co-run mode further examine cases with heavy system workloads and
additional inter-workflow resource contention.
The test scripts are provided by \emph{FaaSFlow}, but for tests with \emph{BeeFlow},
\emph{FaaSFlow}-specific operations (e.g., clearing CouchDB) are removed,
while other parts remain unchanged.

\subsection{Implementation}

We have implemented \emph{BeeFlow} using Golang, including
a behavior tree engine ($\sim$5k LoC),
a system implementation following the reference architecture design ($\sim$3k LoC), and
a few Python scripts ($\sim$0.5k LoC) for adapting the workflows shipped by \emph{FaaSFlow}.

For the message broker, we use NATS \cite{WEB-nats} due to its lightweight and portable design, and
for the storage service, we choose MinIO \cite{WEB-minio} as it is S3-compatible and
widely adopted in self-hosted Serverless systems.
We additionally deploy Prometheus \cite{WEB-prometheus} and Jaeger \cite{WEB-jaeger}
for monitoring workflows.
Other software selections are kept the same as \emph{FaaSFlow} to ensure compatibility.

\subsection{Deployment}

The specifications of the edge testbed have been shown in Table~\ref{tab:firefly} previously. 
And for the cloud testbed, 8 high-profile Aliyun ECS instances are involved,
including 1 \texttt{ecs.g6e.4xlarge} instance and 7 \texttt{ecs.g7.2xlarge} instances.
The detailed specifications are shown in Table~\ref{tab:aliyun}.
For a more comprehensive description of the edge and cloud testbeds,
as well as an auxiliary evaluation in a conventional edge environment,
please refer to \ref{appendix:edge-con}.

\begin{table}[!ht]
    \centering
    \caption{Specifications of Aliyun ECS instances.}
    \vspace{-0.5em}
    \label{tab:aliyun}
    \begin{tabular}{@{}R{2.5cm}|L{5.5cm}@{}}
        \toprule
        \textbf{1*storage node \emph{\footnotesize (\tFaaSFlow only)}}
        & Intel Xeon Platinum 8269CY,
        \mbox{ecs.g6e.4xlarge with 16 vCPU}, \mbox{64GB RAM}, 100GB SSD(3000 IOPS)
        \\\midrule
        \textbf{7*worker node}
        & Intel Xeon Platinum 8369B, 
        \mbox{ecs.g7.2xlarge with 8 vCPU}, \mbox{32GB RAM}, 100GB SSD(3000 IOPS)
        \\\bottomrule
    \end{tabular}
\end{table}

For \emph{FaaSFlow} in the cloud testbed,
the \texttt{ecs.g6e.4xlarge} instance serves both the storage component (\emph{FaaStore}) and
the monitoring stack for auditing container usage, while the 7 \texttt{ecs.g7.2xlarge} instances
serve as worker nodes.
For \emph{BeeFlow} in the cloud testbed, we deploy a separate MinIO instance in each worker
node. Since there is no centralized storage service, the memory capacity in each node is ample, and
the gateway process as well as the message broker
are fairly lightweight, the powerful \texttt{ecs.g6e.4xlarge} instance is not utilized and all of
the gateway process, message broker, and monitoring stack are co-located in a random worker
node.

For \emph{FaaSFlow} in the edge testbed,
the deployment scheme follows the same as described in Section~\ref{sec:motivation-study}.
For \emph{BeeFlow} in the edge testbed,
the main SBC serves all of the gateway process, message broker, and monitoring stack,
and a separate MinIO instance is deployed in each worker node.

\subsection{Performance comparison}

We evaluate the end-to-end performance as described in Section~\ref{sec:benchmarking}.
For the edge testbed, since testing in the co-run mode with \emph{FaaSFlow} is
destructive (leading to Out-Of-Memory errors due to resource shortage), we compare
\emph{BeeFlow} in both modes with \emph{FaaSFlow} in the single mode.
For the cloud testbed, we run \emph{FaaSFlow} and \emph{BeeFlow} in both modes and
compare their results respectively%
\footnote{
    Note that our test results with \emph{FaaSFlow} in the cloud testbed are consistent with
    the results reported by the authors,
    with an exception for the FP workflow, where we use the version that performs real computations
    rather than the simulated version \emph{FaaSFlow} used in their original paper.
}.

Moreover, we record the \emph{function} placement schemes of \emph{FaaSFlow} during tests in the
cloud testbed and replay the exact schemes with \emph{BeeFlow} to distinguish performance
impacts originating from the architecture design and the behavior tree-based scheduling.
We denote this hybrid deployment as \emph{FaaSFlow+arch}.
We do not perform the same deployment in the edge testbed because \emph{FaaSFlow} constantly
evicts \emph{functions} in the edge testbed, making it impossible to mock the same placement schemes
with a different architecture (on the contrary, the \emph{function} placement of \emph{FaaSFlow} remains
stable during all tests in the cloud testbed).

Finally, we note that the per-invocation time limit as well as the per-workflow test time limit
in the scripts are enlarged twice for the edge testbed. These empirical adjustments allow the
testing scripts to render effective and stable benchmarking results for different methods in
the edge testbed.

\textbf{End-to-end speedup.}
The end-to-end speedup of \emph{BeeFlow} is evident in the edge testbed,
as shown in Figure~\ref{fig:eval:e2e-edge},
with \emph{BeeFlow} outperforming \emph{FaaSFlow} in all workflows in both modes
(up to $3.2\times$ in the single mode and $2.7\times$ in the co-run mode).
Even for the Cyc workflow that \emph{FaaSFlow} fails to pass,
\emph{BeeFlow} can run smoothly in both modes.

\begin{figure}[!ht]
    \centering
    \includegraphics[scale=0.75]{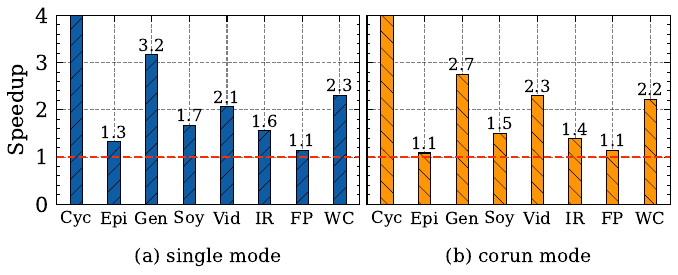}
    \vspace{-0.5em}
    \caption{
        Speedup over \emph{FaaSFlow} in the single mode (edge).
        \emph{%
            The Cyc workflow is disruptive in the edge testbed with FaaSFlow,
            while it runs smoothly with BeeFlow, and so the speedup is infinite.
        }
    }
    \label{fig:eval:e2e-edge}
\end{figure}

Figure~\ref{fig:eval:e2e-cloud} displays the speedup over \emph{FaaSFlow} achieved by
\emph{FaaSFlow+arch} and \emph{BeeFlow} in the cloud testbed.
The results show that \emph{FaaSFlow+arch}, which merely replaces the architecture design of
\emph{FaaSFlow}, cannot always surpass \emph{FaaSFlow}. Nonetheless, it still achieves improvements
in some workflows (e.g., Gen and WC), suggesting the benefits
gained from mitigating the single-point bottleneck of the storage service.
Further with contention-aware partitioning and placement, \emph{BeeFlow} achieves
un-compromised performance in all workflows even in the absence of in-memory caching.

\begin{figure}[!ht]
    \centering
    \includegraphics[scale=0.75]{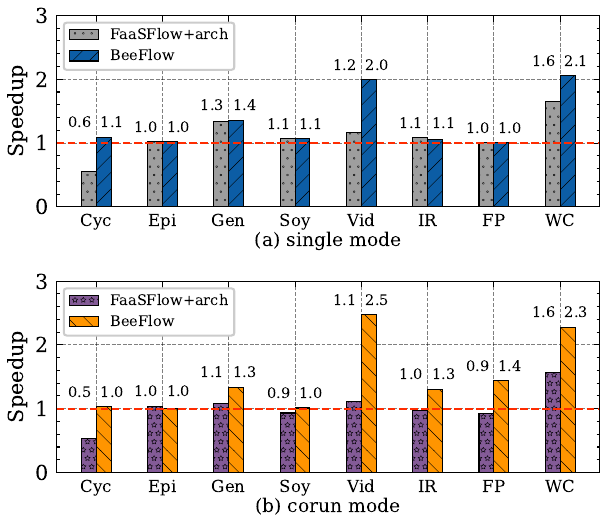}
    \vspace{-0.5em}
    \caption{
        Speedup over \emph{FaaSFlow} in both modes, respectively (cloud).
    }
    \label{fig:eval:e2e-cloud}
\end{figure}

\textbf{Factor analysis.}
Since there are no real computations involved in the 4 scientific workflows, the only factor that
may impact the end-to-end performance is I/O, mostly network I/O.
The 4 real-world application workflows further help examine the cases with both computation and
communication factors.

To better connect the performance results with different methods and workflows,
we estimate the initialization delay, input fetching delay, execution delay, output
putting delay, and other dynamicity for each workflow \emph{function} through runtime traces, and
then align these \emph{functions} based on the workflow definitions to create Gantt charts.
Specifically, a Gantt chart spans the full execution of a workflow by its horizontal axis,
with each horizontal lane representing a workflow \emph{function}, where occupied areas in the lane
stand for different periods of the \emph{function} relative to the workflow execution.

Firstly, the Gantt chart for the Cyc workflow in Figure~\ref{fig:gantt:cyc} shows that the Cyc
workflow generates massive I/O transfer and contention, which explains the test failures with
\emph{FaaSFlow} in the edge testbed.
As for Cyc in the cloud testbed, \emph{FaaSFlow+arch} is only 0.6$\times$ fast as \emph{FaaSFlow}
since in-memory caching is not exploited to accelerate the massive transfer.
However, with contention-aware scheduling, \emph{BeeFlow} lifts up the performance of
\emph{FaaSFlow+arch} and slightly outperforms \emph{FaaSFlow}.

The Gen workflow instead generates less transfer but severe I/O contention, as shown in
Figure~\ref{fig:gantt:gen}. In this case, \emph{FaaSFlow+arch} and \emph{BeeFlow} achieve similar
speedup (${\sim}1.3\times$) in the cloud testbed, suggesting the centralized storage service
of \emph{FaaSFlow} is to blame. 

For the remaining Epi and Soy workflows shown in Figure~\ref{fig:gantt:epi} and
Figure~\ref{fig:gantt:soy}, there are much fewer I/O transfer and contention, therefore
all of \emph{FaaSFlow}, \emph{FaaSFlow+arch}, and \emph{BeeFlow} achieve similar performance
in the cloud testbed.
However, for the edge testbed, even these cases with moderate I/O transfer and contention can
still result in observable impacts, where \emph{BeeFlow} achieves remarkable speedups ($1.3\times$
and $1.7\times$ for Epi and Soy, respectively), proving the effectiveness of \emph{BeeFlow} for
resource-constrained edge nodes.

\begin{figure}[!ht]
    \centering
    \begin{subfigure}[b]{0.45\linewidth}
        \centering
        \includegraphics[scale=0.34]{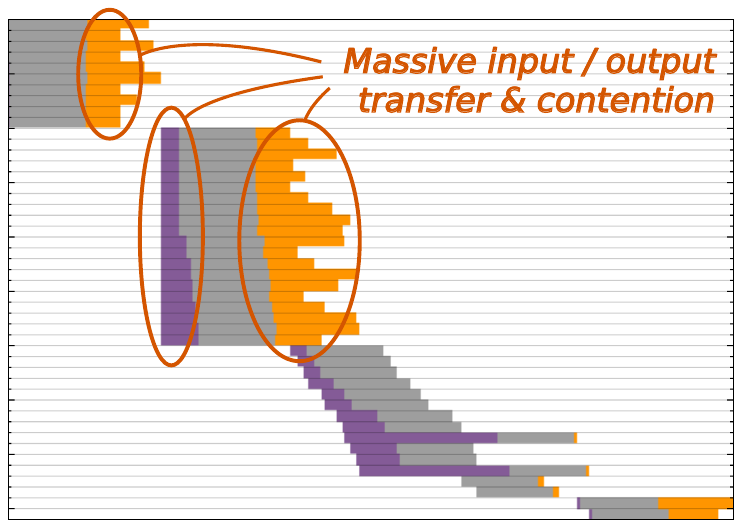}
        \caption{Cyc (speedup: 0.6/1.1/$\infty$)}
        \label{fig:gantt:cyc}
    \end{subfigure}
    \hfill
    \begin{subfigure}[b]{0.54\linewidth}
        \centering
        \includegraphics[scale=0.34,right]{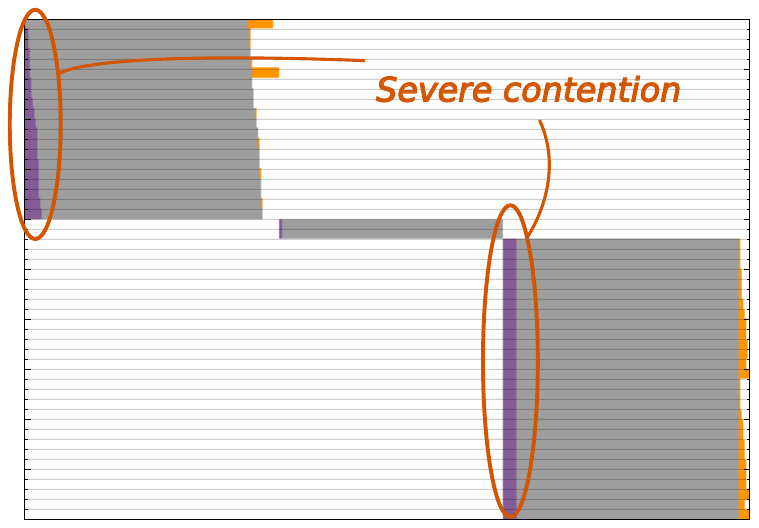}
        \caption{Gen (speedup: 1.3/1.4/3.2)}
        \label{fig:gantt:gen}
    \end{subfigure}
    \hfill
    \begin{subfigure}[b]{0.45\linewidth}
        \centering
        \includegraphics[scale=0.34]{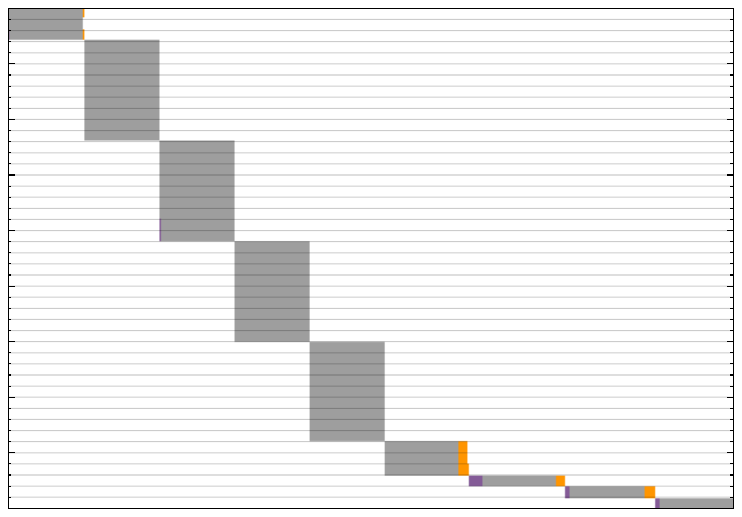}
        \caption{Epi (speedup: 1.0/1.0/1.3)}
        \label{fig:gantt:epi}
    \end{subfigure}
    \hfill
    \begin{subfigure}[b]{0.54\linewidth}
        \includegraphics[scale=0.34,right]{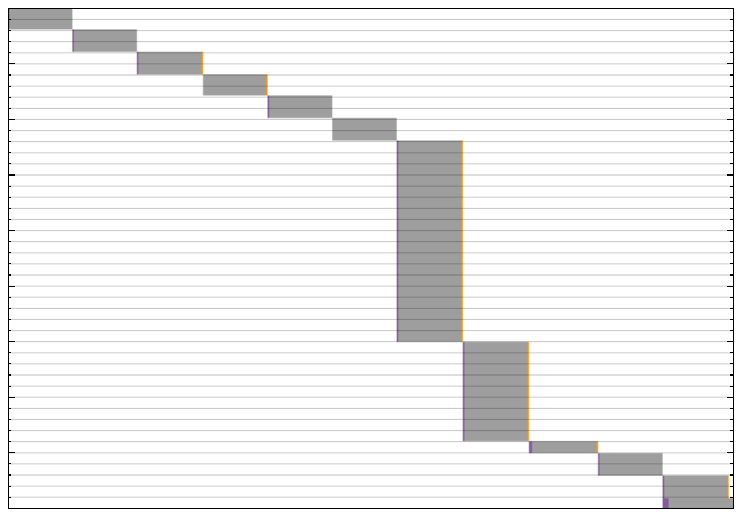}
        \caption{Soy (speedup: 1.1/1.1/1.7)}
        \label{fig:gantt:soy}
    \end{subfigure}
    \vspace{-0.5em}
    \caption{Gantt charts for scientific workflows.
        \emph{%
            Periods of function initialization, input fetching, execution, and output putting
            are colored with gray, violet, light gray, and yellow, respectively
            (better viewed in color).
            Speedups of FaaSFlow+arch (cloud), BeeFlow (cloud), and BeeFlow (edge) in the single mode
            are annotated in each sub-caption for easier comparison.
            These conventions remain the same for the following Gantt charts.
        }
    }
    \label{fig:gantt:sci}
\end{figure}

Finally, we look at two of the application workflows: Vid in Figure~\ref{fig:gantt:vid}
and WC in Figure~\ref{fig:gantt:wc}, which both use the \emph{foreach} logic%
\footnote{
    The \emph{foreach} logic can be regarded as a dynamic variant of the \emph{parallel} composite
    in the behavior tree model and has been realized in our implementation.
}
in their workflow definitions.
While \emph{FaaSFlow} intends to co-locate the \emph{foreach} branches to exploit data locality
of intermediate data, which exacerbates both computation and communication contention,
\emph{BeeFlow} is more conservative by both mitigating and balancing contention,
thus achieving significant speedups in both the edge and cloud testbeds.

\begin{figure}[!ht]
    \centering
    \begin{subfigure}[b]{0.49\linewidth}
        \centering
        \includegraphics[scale=0.35]{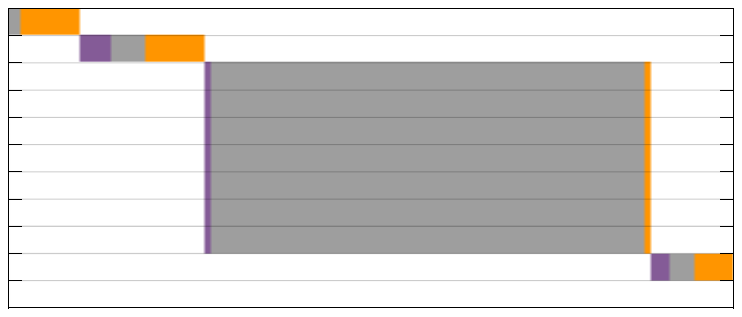}
        \caption{Vid (speedup: 1.2/2.0/2.1)}
        \label{fig:gantt:vid}
    \end{subfigure}
    \hfill
    \begin{subfigure}[b]{0.49\linewidth}
        \centering
        \includegraphics[scale=0.35]{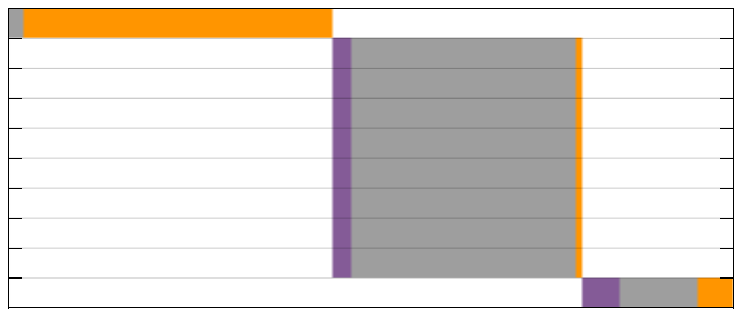}
        \caption{WC (speedup: 1.6/2.1/2.3)}
        \label{fig:gantt:wc}
    \end{subfigure}
    \vspace{-0.5em}
    \caption{Gantt charts for Vid and WC.}
    \label{fig:gantt:vid-wc}
\end{figure}

\textbf{Robustness and workload balance.}
In Figure~\ref{fig:eval:e2e-latency}, we normalize the latency results over \emph{FaaSFlow} in
the single mode for both the edge and cloud testbeds. The results show that
\emph{BeeFlow} achieves consistent performance in both modes, suggesting \emph{BeeFlow} is
more robust to heavy system workloads and mutual interference among workflows.

\begin{figure}[!ht]
    \centering
    \includegraphics[scale=0.75]{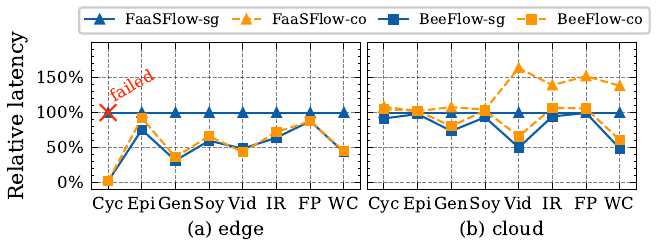}
    \vspace{-0.5em}
    \caption{
        Relative latency w.r.t. \emph{FaaSFlow} in the single mode.
    }
    \label{fig:eval:e2e-latency}
\end{figure}

\begin{figure}[!ht]
    \centering
    \includegraphics[width=0.85\linewidth]{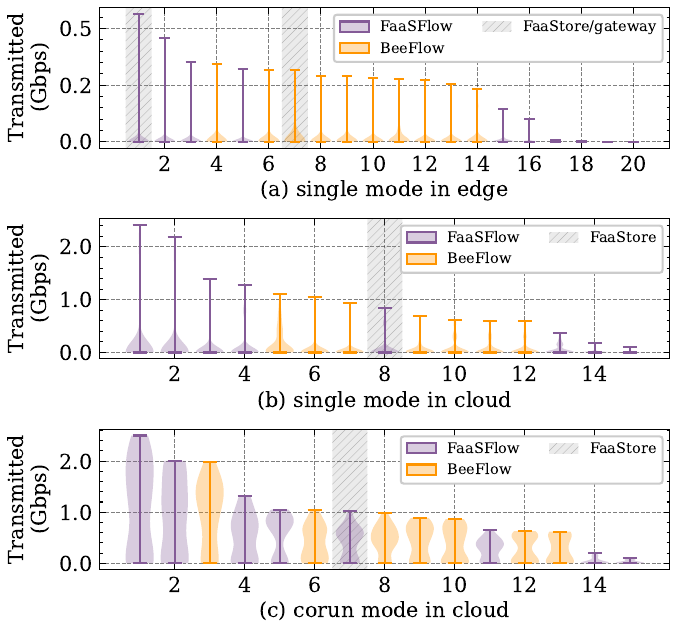}
    \vspace{-0.5em}
    \caption{
        Per-SBC distribution of 5-second averaged transmitted speeds.
        \emph{%
            The Cyc workflow is excluded for FaaSFlow in the edge testbed,
            the violin plots are sorted w.r.t. the maximum speed,
            and the node serving as the storage node (FaaStore) or the gateway node
            is marked with a gray shadow.
        }
    }
    \label{fig:eval:trans}
\end{figure}

Lastly, Figure~\ref{fig:eval:trans} compares the transmission speed distributions between
\emph{BeeFlow} and \emph{FaaSFlow}, reflecting that \emph{BeeFlow} evenly distributes
data transmission across nodes, in contrast to \emph{FaaSFlow} with significant disparity.
The consistency in the latency reduction and the balanced data transmission again validates the
effectiveness of \emph{BeeFlow} in mitigating intra-workflow resource contention
and balancing workloads across nodes.

}
\section{Implications and Limitations}{%
\textbf{Resource contention can lead to significant performance degradation in
resource-constrained edge clusters.} Existing resource management platforms
focus primarily on reservation-based resource allocation, but they often overlook
transient resource contention.
Since resources per-SoC/SBC are generally more limited than those in conventional
servers, transient resource contention can introduce significant negative
impacts on resource-constrained SoC/SBC-clusters.
\emph{BeeFlow} addresses the
resource contention problem by modeling workflows using behavior trees, generating
contention-free function collections as resource control units, and employing
runtime traces-instrumented static analysis to derive concurrent invocation awareness.

\textbf{Centralized components in distributed systems should be carefully re-examined in
resource-constrained edge clusters.} Given the limited resources per-SoC/SBC,
even moderately stressed centralized components may slowdown the whole application.
\emph{BeeFlow} proposes a reference architecture design that mitigates centralized components
and remains minimal to facilitate easy integration with other systems in edge computing.

\textbf{Limitations.} There are several aspects in which our experimental
study can be improved:\begin{itemize}
    \item Our experimental study currently uses 4 popular scientific workflows and 4 representative
        application workflows provided by \emph{FaaSFlow} as the benchmark. Although
        these workflows offer a diverse range, they are not specifically tailored for
        SoC/SBC-clusters. We plan to explore more applications that can fully
        harness the power of SoC/SBC-clusters. 
    \item The adoption of SoC/SBC-clusters is still relatively uncommon, and we have only evaluated
        \emph{BeeFlow} in a single SBC-cluster configuration so far. We plan to investigate
        different forms of SoC/SBC-clusters further.
\end{itemize}

}
\section{Related works}{%
\label{sec:related}
\textbf{Serverless workflow optimization.}
Presently, most research on Serverless workflow optimization
\cite{DBLP-cs19-survey, DBLP-atc21-sonic, DBLP-atc18-sand} is cloud-oriented and
assumes DAGs or DAG variants, such as
Wukong \cite{DBLP-socc20-wukong},
Xanadu \cite{DBLP-middleware20-xanadu},
Atoll \cite{DBLP-socc21-atoll}, and
FaaSFlow \cite{DBLP-asplos22-FaaSFlow}.
Wukong proposes re-purposing Lambda executors to exploit data locality through DAG-based static
and dynamic scheduling.
Xanadu implements speculative and just-in-time provisioning by estimating the
most-likely-path from DAGs to alleviate cascading cold starts in function chains.
Atoll leverages chain and DAG information to achieve deadline-aware scheduling, while
FaaSFlow adopts a worker-side workflow schedule pattern and applies the critical path analysis
to figure out effective schemes for in-memory caching-based state sharing.
Recent studies have also begun to evaluate Serverless workflows at the edge.
DPE \cite{DBLP:journals/tpds/DengZXZJLYDZ22} investigates proactive stream mapping and data
splitting in DAGs for Serverless edge computing.
Cicconetti \emph{et al.} \cite{DBLP:journals/percom/CicconettiCP22} suggest applying data locality
principles to address low link speed issues in edge networks for Serverless workflows.
Poojara \emph{et al.} \cite{DBLP:journals/fgcs/PoojaraDJS22} introduce a layered edge-fog-cloud
Serverless data pipeline for IoT data processing. 

Distinctly, our work proposes modeling Serverless workflows using behavior trees, which enables
convenient synthesis and analysis while retaining full workflow expressiveness. Additionally,
our work specially focuses on accommodating the resource-constrained nature of edge clusters
through contention-aware workflow scheduling.

\textbf{Serverless edge computing.}
With the growing demand for intelligent applications at the network edge
\cite{DBLP:journals/pieee/ZhouCLZLZ19,10128754}, the elastic Serverless architecture is
increasingly being explored in edge computing
\cite{DBLP:conf/acsw/AslanpourTCJSTA21,9931467}.
Although existing Serverless ecosystems are primarily designed for cloud computing,
continuous efforts are being made to develop edge-native Serverless computing systems.
WOW \cite{DBLP:conf/ccgrid/GackstatterFD22} evaluates WebAssembly execution in Apache OpenWhisk
to reduce cold-start latency, optimizes memory consumption, and increases function throughput
on low-end edge computing equipment.
Sledge \cite{DBLP:conf/eurosys/LyuCAP022} extends WebAssembly-based Serverless runtime for efficient
DAG support at the edge.
Zheng \emph{et al.} \cite{ZHENG2023105}
investigate the dependency package caching problem in Serverless edge computing. 

Given the minimal architecture design of \emph{BeeFlow}, integrating our work into these
edge-native systems should be straightforward and is considered as an interesting direction
for future work.

}
\section{Conclusion}{%
\label{sec:conclusion}

In this work, we propose \emph{BeeFlow}, a \ul{b}ehavior tr\ul{ee}-based Serverless work\ul{flow}
system.
It utilizes behavior trees for Serverless workflow modeling, based on which contention-free
\emph{function} collections and awareness of exact and empirical concurrent invocations are
derived.
It then leverages the \emph{function} collections as resource control units and employs
contention-aware workflow partitioning and placement algorithms to distribute \emph{function}
collections across server nodes, which mitigates intra-workflow resource contention and
balances both intra- and inter-workflow resource contention.
Evaluation results based on a system prototype in two real-world testbeds confirm that
\emph{BeeFlow} achieves significant speedup and effectively accommodates the resource-constrained
nature of edge clusters.

This work lays the foundation for a novel approach to modeling and scheduling Serverless
workflows. In future research, we aim to explore \emph{BeeFlow} in more complex edge scenarios,
particularly in device-edge-cloud-based layered architectures, where the comprehensive hierarchy
and modularity of behavior trees will facilitate intelligent, dynamic, and seamless collaboration
throughout the computing continuum.

}
\section*{Acknowledgements}{%
We would like to thank the anonymous reviewers for their insightful and constructive feedback.
This work was supported in part by the National Science Foundation of China (No. 61972432);
the Program for Guangdong Introducing Innovative and Entrepreneurial Teams (No.2017ZT07X355).
}
\bibliographystyle{elsarticle-num} 
\bibliography{related.bib}

\begin{thebibliography}{10}
\expandafter\ifx\csname url\endcsname\relax
  \def\url#1{\texttt{#1}}\fi
\expandafter\ifx\csname urlprefix\endcsname\relax\def\urlprefix{URL }\fi
\expandafter\ifx\csname href\endcsname\relax
  \def\href#1#2{#2} \def\path#1{#1}\fi

\bibitem{DBLP-arxiv19-berkeley}
E.~Jonas, J.~Schleier{-}Smith, V.~Sreekanti, C.~Tsai, A.~Khandelwal, Q.~Pu,
  V.~Shankar, J.~Carreira, K.~Krauth, N.~J. Yadwadkar, J.~E. Gonzalez, R.~A.
  Popa, I.~Stoica, D.~A. Patterson,
  \href{http://arxiv.org/abs/1902.03383}{Cloud programming simplified: {A}
  berkeley view on serverless computing}, CoRR abs/1902.03383 (2019).
\newblock \href {http://arxiv.org/abs/1902.03383} {\path{arXiv:1902.03383}}.
\newline\urlprefix\url{http://arxiv.org/abs/1902.03383}

\bibitem{WEB-Amazon-Lambda}
Amazon, \href{https://aws.amazon.com/lambda}{{AWS Lambda}}, accessed:
  2023-06-18.
\newline\urlprefix\url{https://aws.amazon.com/lambda}

\bibitem{WEB-Google-Functions}
Google, \href{https://cloud.google.com/functions}{{Cloud Functions}}, accessed:
  2023-06-18.
\newline\urlprefix\url{https://cloud.google.com/functions}

\bibitem{WEB-Microsoft-Functions}
Microsoft, \href{https://azure.microsoft.com/services/functions}{{Azure
  Functions}}, accessed: 2023-06-18.
\newline\urlprefix\url{https://azure.microsoft.com/services/functions}

\bibitem{WEB-Alibaba-FC}
{Alibaba Cloud},
  \href{https://alibabacloud.com/product/function-compute}{{Function Compute}},
  accessed: 2023-06-18.
\newline\urlprefix\url{https://alibabacloud.com/product/function-compute}

\bibitem{WEB-Apache-OpenWhisk}
{Apache Openwhisk}, \href{https://openwhisk.apache.org}{Open source serverless
  cloud platform}, accessed: 2023-06-18.
\newline\urlprefix\url{https://openwhisk.apache.org}

\bibitem{WEB-OpenFaaS}
OpenFaaS, \href{https://www.openfaas.com}{Serverless functions, made simple},
  accessed: 2023-06-18.
\newline\urlprefix\url{https://www.openfaas.com}

\bibitem{ACM-cs21-sjtu}
Z.~Li, L.~Guo, J.~Cheng, Q.~Chen, B.~He, M.~Guo,
  \href{https://doi.org/10.1145/3508360}{The serverless computing survey: A
  technical primer for design architecture}, ACM Comput. Surv. 54~(10s) (sep
  2022).
\newblock \href {https://doi.org/10.1145/3508360} {\path{doi:10.1145/3508360}}.
\newline\urlprefix\url{https://doi.org/10.1145/3508360}

\bibitem{DBLP-jcloudc21-survey}
H.~B. Hassan, S.~A. Barakat, Q.~I. Sarhan,
  \href{https://doi.org/10.1186/s13677-021-00253-7}{Survey on serverless
  computing}, J. Cloud Comput. 10~(1) (2021) 39.
\newblock \href {https://doi.org/10.1186/s13677-021-00253-7}
  {\path{doi:10.1186/s13677-021-00253-7}}.
\newline\urlprefix\url{https://doi.org/10.1186/s13677-021-00253-7}

\bibitem{DBLP-cacm21-survey}
J.~Schleier{-}Smith, V.~Sreekanti, A.~Khandelwal, J.~Carreira, N.~J. Yadwadkar,
  R.~A. Popa, J.~E. Gonzalez, I.~Stoica, D.~A. Patterson,
  \href{https://doi.org/10.1145/3406011}{What serverless computing is and
  should become: the next phase of cloud computing}, Commun. {ACM} 64~(5)
  (2021) 76--84.
\newblock \href {https://doi.org/10.1145/3406011} {\path{doi:10.1145/3406011}}.
\newline\urlprefix\url{https://doi.org/10.1145/3406011}

\bibitem{WEB-Amazon-ASL}
Amazon, \href{https://states-language.net}{{Amazon States Language}}, accessed:
  2023-06-18.
\newline\urlprefix\url{https://states-language.net}

\bibitem{DBLP-socc20-wukong}
B.~Carver, J.~Zhang, A.~Wang, A.~Anwar, P.~Wu, Y.~Cheng,
  \href{https://doi.org/10.1145/3419111.3421286}{Wukong: a scalable and
  locality-enhanced framework for serverless parallel computing}, in:
  R.~Fonseca, C.~Delimitrou, B.~C. Ooi (Eds.), SoCC '20: {ACM} Symposium on
  Cloud Computing, Virtual Event, USA, October 19-21, 2020, {ACM}, 2020, pp.
  1--15.
\newblock \href {https://doi.org/10.1145/3419111.3421286}
  {\path{doi:10.1145/3419111.3421286}}.
\newline\urlprefix\url{https://doi.org/10.1145/3419111.3421286}

\bibitem{DBLP-socc21-atoll}
A.~Singhvi, A.~Balasubramanian, K.~Houck, M.~D. Shaikh, S.~Venkataraman,
  A.~Akella, \href{https://doi.org/10.1145/3472883.3486981}{Atoll: {A} scalable
  low-latency serverless platform}, in: C.~Curino, G.~Koutrika, R.~Netravali
  (Eds.), SoCC '21: {ACM} Symposium on Cloud Computing, Seattle, WA, USA,
  November 1 - 4, 2021, {ACM}, 2021, pp. 138--152.
\newblock \href {https://doi.org/10.1145/3472883.3486981}
  {\path{doi:10.1145/3472883.3486981}}.
\newline\urlprefix\url{https://doi.org/10.1145/3472883.3486981}

\bibitem{DBLP-asplos22-FaaSFlow}
Z.~Li, Y.~Liu, L.~Guo, Q.~Chen, J.~Cheng, W.~Zheng, M.~Guo,
  \href{https://doi.org/10.1145/3503222.3507717}{Faasflow: enable efficient
  workflow execution for function-as-a-service}, in: B.~Falsafi, M.~Ferdman,
  S.~Lu, T.~F. Wenisch (Eds.), {ASPLOS} '22: 27th {ACM} International
  Conference on Architectural Support for Programming Languages and Operating
  Systems, Lausanne, Switzerland, 28 February 2022 - 4 March 2022, {ACM}, 2022,
  pp. 782--796.
\newblock \href {https://doi.org/10.1145/3503222.3507717}
  {\path{doi:10.1145/3503222.3507717}}.
\newline\urlprefix\url{https://doi.org/10.1145/3503222.3507717}

\bibitem{10163866}
L.~Zhao, E.~Zhang, S.~Wan, A.~Hawbani, A.~Y. Al-Dubai, G.~Min, A.~Y. Zomaya,
  Meson: A mobility-aware dependent task offloading scheme for urban vehicular
  edge computing, IEEE Transactions on Mobile Computing~(01) (5555) 1--15.
\newblock \href {https://doi.org/10.1109/TMC.2023.3289611}
  {\path{doi:10.1109/TMC.2023.3289611}}.

\bibitem{10128791}
L.~Wang, X.~Deng, J.~Gui, X.~Chen, S.~Wan, Microservice-oriented service
  placement for mobile edge computing in sustainable internet of vehicles, IEEE
  Transactions on Intelligent Transportation Systems (2023) 1--15\href
  {https://doi.org/10.1109/TITS.2023.3274307}
  {\path{doi:10.1109/TITS.2023.3274307}}.

\bibitem{DBLP:conf/edge/RajputKCWK22}
K.~R. Rajput, C.~D. Kulkarni, B.~Cho, W.~Wang, I.~K. Kim,
  \href{https://doi.org/10.1109/EDGE55608.2022.00024}{Edgefaasbench:
  Benchmarking edge devices using serverless computing}, in: C.~A. Ardagna,
  H.~Bian, C.~K. Chang, R.~N. Chang, E.~Damiani, G.~Elia, Q.~He, V.~Puig,
  R.~Ward, F.~Xhafa, J.~Zhang (Eds.), {IEEE} International Conference on Edge
  Computing and Communications, {EDGE} 2022, Barcelona, Spain, July 10-16,
  2022, {IEEE}, 2022, pp. 93--103.
\newblock \href {https://doi.org/10.1109/EDGE55608.2022.00024}
  {\path{doi:10.1109/EDGE55608.2022.00024}}.
\newline\urlprefix\url{https://doi.org/10.1109/EDGE55608.2022.00024}

\bibitem{DBLP:conf/eurosys/LyuCAP022}
X.~Lyu, L.~Cherkasova, R.~C. Aitken, G.~Parmer, T.~Wood,
  \href{https://doi.org/10.1145/3517206.3526274}{Towards efficient processing
  of latency-sensitive serverless dags at the edge}, in: A.~Y. Ding, V.~Hilt
  (Eds.), EdgeSys@EuroSys 2022: Proceedings of the 5th International Workshop
  on Edge Systems, Analytics and Networking, Rennes, France, April 5 - 8, 2022,
  {ACM}, 2022, pp. 49--54.
\newblock \href {https://doi.org/10.1145/3517206.3526274}
  {\path{doi:10.1145/3517206.3526274}}.
\newline\urlprefix\url{https://doi.org/10.1145/3517206.3526274}

\bibitem{DBLP:conf/ieeesec/XuZW22}
M.~Xu, L.~Zhang, S.~Wang,
  \href{https://doi.org/10.1109/SEC54971.2022.00024}{Position paper: Renovating
  edge servers with {ARM} socs}, in: 7th {IEEE/ACM} Symposium on Edge
  Computing, {SEC} 2022, Seattle, WA, USA, December 5-8, 2022, {IEEE}, 2022,
  pp. 216--223.
\newblock \href {https://doi.org/10.1109/SEC54971.2022.00024}
  {\path{doi:10.1109/SEC54971.2022.00024}}.
\newline\urlprefix\url{https://doi.org/10.1109/SEC54971.2022.00024}

\bibitem{DBLP:journals/corr/abs-2212-12842}
L.~Zhang, Z.~Fu, B.~Shi, X.~Li, R.~Lai, C.~Chen, A.~Zhou, X.~Ma, S.~Wang,
  M.~Xu, \href{https://doi.org/10.48550/arXiv.2212.12842}{Soc-cluster as an
  edge server: an application-driven measurement study}, CoRR abs/2212.12842
  (2022).
\newblock \href {http://arxiv.org/abs/2212.12842} {\path{arXiv:2212.12842}},
  \href {https://doi.org/10.48550/arXiv.2212.12842}
  {\path{doi:10.48550/arXiv.2212.12842}}.
\newline\urlprefix\url{https://doi.org/10.48550/arXiv.2212.12842}

\bibitem{DBLP-ppopp22-Mashup}
R.~B. Roy, T.~Patel, V.~Gadepally, D.~Tiwari,
  \href{https://doi.org/10.1145/3503221.3508407}{Mashup: making serverless
  computing useful for {HPC} workflows via hybrid execution}, in: J.~Lee,
  K.~Agrawal, M.~F. Spear (Eds.), PPoPP '22: 27th {ACM} {SIGPLAN} Symposium on
  Principles and Practice of Parallel Programming, Seoul, Republic of Korea,
  April 2 - 6, 2022, {ACM}, 2022, pp. 46--60.
\newblock \href {https://doi.org/10.1145/3503221.3508407}
  {\path{doi:10.1145/3503221.3508407}}.
\newline\urlprefix\url{https://doi.org/10.1145/3503221.3508407}

\bibitem{DBLP-socc18-sprocket}
L.~Ao, L.~Izhikevich, G.~M. Voelker, G.~Porter,
  \href{https://doi.org/10.1145/3267809.3267815}{Sprocket: {A} serverless video
  processing framework}, in: Proceedings of the {ACM} Symposium on Cloud
  Computing, SoCC 2018, Carlsbad, CA, USA, October 11-13, 2018, {ACM}, 2018,
  pp. 263--274.
\newblock \href {https://doi.org/10.1145/3267809.3267815}
  {\path{doi:10.1145/3267809.3267815}}.
\newline\urlprefix\url{https://doi.org/10.1145/3267809.3267815}

\bibitem{DBLP-ras22-survey}
M.~Iovino, E.~Scukins, J.~Styrud, P.~{\"{O}}gren, C.~Smith,
  \href{https://doi.org/10.1016/j.robot.2022.104096}{A survey of behavior trees
  in robotics and {AI}}, Robotics Auton. Syst. 154 (2022) 104096.
\newblock \href {https://doi.org/10.1016/j.robot.2022.104096}
  {\path{doi:10.1016/j.robot.2022.104096}}.
\newline\urlprefix\url{https://doi.org/10.1016/j.robot.2022.104096}

\bibitem{WEB-Amazon-Step}
Amazon, \href{https://aws.amazon.com/step-functions}{{AWS} step functions},
  accessed: 2023-06-18.
\newline\urlprefix\url{https://aws.amazon.com/step-functions}

\bibitem{WEB-Google-workflows}
Google, \href{https://cloud.google.com/workflows}{Google cloud workflows},
  accessed: 2023-06-18.
\newline\urlprefix\url{https://cloud.google.com/workflows}

\bibitem{WEB-Microsoft-logic}
Microsoft, \href{https://azure.microsoft.com/services/logic-apps}{Azure logic
  apps}, accessed: 2023-06-18.
\newline\urlprefix\url{https://azure.microsoft.com/services/logic-apps}

\bibitem{WEB-CNCF-workflow}
CNCF, \href{https://serverlessworkflow.github.io}{Serverless workflow},
  accessed: 2023-06-18.
\newline\urlprefix\url{https://serverlessworkflow.github.io}

\bibitem{WEB-cadence}
Cadence, \href{https://cadenceworkflow.io/}{Fault-tolerant stateful code
  platform}, accessed: 2023-06-18.
\newline\urlprefix\url{https://cadenceworkflow.io/}

\bibitem{WEB-temporal}
Temporal, \href{https://temporal.io/}{Open source durable execution platform},
  accessed: 2023-06-18.
\newline\urlprefix\url{https://temporal.io/}

\bibitem{WEB-swf-lambda}
{AWS SWF}, \href{https://go.aws/41jdolK}{{AWS} lambda tasks}, accessed:
  2023-06-18.
\newline\urlprefix\url{https://go.aws/41jdolK}

\bibitem{DBLP-atc18-sand}
I.~E. Akkus, R.~Chen, I.~Rimac, M.~Stein, K.~Satzke, A.~Beck, P.~Aditya,
  V.~Hilt, {SAND:} towards high-performance serverless computing, in: H.~S.
  Gunawi, B.~C. Reed (Eds.), 2018 {USENIX} Annual Technical Conference,
  {USENIX} {ATC} 2018, Boston, MA, USA, July 11-13, 2018, {USENIX} Association,
  2018, pp. 923--935.

\bibitem{DBLP-cs19-survey}
M.~Adhikari, T.~Amgoth, S.~N. Srirama, \href{https://doi.org/10.1145/3325097}{A
  survey on scheduling strategies for workflows in cloud environment and
  emerging trends}, {ACM} Comput. Surv. 52~(4) (2019) 68:1--68:36.
\newblock \href {https://doi.org/10.1145/3325097} {\path{doi:10.1145/3325097}}.
\newline\urlprefix\url{https://doi.org/10.1145/3325097}

\bibitem{DBLP-atc21-sonic}
A.~Mahgoub, K.~Shankar, S.~Mitra, A.~Klimovic, S.~Chaterji, S.~Bagchi, {SONIC:}
  application-aware data passing for chained serverless applications, in:
  I.~Calciu, G.~Kuenning (Eds.), 2021 {USENIX} Annual Technical Conference,
  {USENIX} {ATC} 2021, July 14-16, 2021, {USENIX} Association, 2021, pp.
  285--301.

\bibitem{DBLP-middleware20-xanadu}
N.~Daw, U.~Bellur, P.~Kulkarni,
  \href{https://doi.org/10.1145/3423211.3425690}{Xanadu: Mitigating cascading
  cold starts in serverless function chain deployments}, in: D.~D. Silva,
  R.~Kapitza (Eds.), Middleware '20: 21st International Middleware Conference,
  Delft, The Netherlands, December 7-11, 2020, {ACM}, 2020, pp. 356--370.
\newblock \href {https://doi.org/10.1145/3423211.3425690}
  {\path{doi:10.1145/3423211.3425690}}.
\newline\urlprefix\url{https://doi.org/10.1145/3423211.3425690}

\bibitem{DBLP:journals/tnsm/WangLSB22}
S.~Wang, X.~Li, Q.~Z. Sheng, A.~Beheshti,
  \href{https://doi.org/10.1109/TNSM.2022.3181145}{Performance analysis and
  optimization on scheduling stochastic cloud service requests: {A} survey},
  {IEEE} Trans. Netw. Serv. Manag. 19~(3) (2022) 3587--3602.
\newblock \href {https://doi.org/10.1109/TNSM.2022.3181145}
  {\path{doi:10.1109/TNSM.2022.3181145}}.
\newline\urlprefix\url{https://doi.org/10.1109/TNSM.2022.3181145}

\bibitem{dcg/systems}
M.~A. Demir, A complete, automated and scalable framework for science and
  engineering, Ph.D. thesis, copyright - Database copyright ProQuest LLC;
  ProQuest does not claim copyright in the individual underlying works; Last
  updated - 2023-06-21 (2020).

\bibitem{DBLP:journals/jcloudc/KraemerWA21}
M.~Kr{\"{a}}mer, H.~M. W{\"{u}}rz, C.~Altenhofen,
  \href{https://doi.org/10.1186/s13677-021-00229-7}{Executing cyclic scientific
  workflows in the cloud}, J. Cloud Comput. 10~(1) (2021) 25.
\newblock \href {https://doi.org/10.1186/s13677-021-00229-7}
  {\path{doi:10.1186/s13677-021-00229-7}}.
\newline\urlprefix\url{https://doi.org/10.1186/s13677-021-00229-7}

\bibitem{DBLP:conf/skg/WuSL11}
R.~Wu, L.~Shuai, H.~Liao, \href{https://doi.org/10.1109/SKG.2011.46}{Cyclic
  workflow execution mechanism on top of mapreduce framework}, in: Seventh
  International Conference on Semantics Knowledge and Grid {(SKG} 2011),
  Beijing, China, October 24-26, 2011, {IEEE} Computer Society, 2011, pp.
  28--35.
\newblock \href {https://doi.org/10.1109/SKG.2011.46}
  {\path{doi:10.1109/SKG.2011.46}}.
\newline\urlprefix\url{https://doi.org/10.1109/SKG.2011.46}

\bibitem{DBLP:journals/fgcs/DeelmanVJRCMMCS15}
E.~Deelman, K.~Vahi, G.~Juve, M.~Rynge, S.~Callaghan, P.~Maechling, R.~Mayani,
  W.~Chen, R.~F. da~Silva, M.~Livny, R.~K. Wenger,
  \href{https://doi.org/10.1016/j.future.2014.10.008}{Pegasus, a workflow
  management system for science automation}, Future Gener. Comput. Syst. 46
  (2015) 17--35.
\newblock \href {https://doi.org/10.1016/j.future.2014.10.008}
  {\path{doi:10.1016/j.future.2014.10.008}}.
\newline\urlprefix\url{https://doi.org/10.1016/j.future.2014.10.008}

\bibitem{WEB-pegasus}
wfcommons, \href{https://github.com/wfcommons/pegasus-instances}{Collection of
  workflow execution instances for the pegasus workflow management system},
  accessed: 2023-06-18.
\newline\urlprefix\url{https://github.com/wfcommons/pegasus-instances}

\bibitem{WEB-firefly}
Firefly,
  \href{https://en.t-firefly.com/product/clusterserver}{{CSR1-N10R3399}},
  accessed: 2023-06-18.
\newline\urlprefix\url{https://en.t-firefly.com/product/clusterserver}

\bibitem{DBLP-icra22-planning}
J.~Styrud, M.~Iovino, M.~Norrl{\"{o}}f, M.~Bj{\"{o}}rkman, C.~Smith,
  \href{https://doi.org/10.1109/ICRA46639.2022.9812086}{Combining planning and
  learning of behavior trees for robotic assembly}, in: 2022 International
  Conference on Robotics and Automation, {ICRA} 2022, Philadelphia, PA, USA,
  May 23-27, 2022, {IEEE}, 2022, pp. 11511--11517.
\newblock \href {https://doi.org/10.1109/ICRA46639.2022.9812086}
  {\path{doi:10.1109/ICRA46639.2022.9812086}}.
\newline\urlprefix\url{https://doi.org/10.1109/ICRA46639.2022.9812086}

\bibitem{DBLP-aaai21-planning}
Z.~Cai, M.~Li, W.~Huang, W.~Yang, {BT} expansion: a sound and complete
  algorithm for behavior planning of intelligent robots with behavior trees,
  in: Thirty-Fifth {AAAI} Conference on Artificial Intelligence, {AAAI} 2021,
  Thirty-Third Conference on Innovative Applications of Artificial
  Intelligence, {IAAI} 2021, The Eleventh Symposium on Educational Advances in
  Artificial Intelligence, {EAAI} 2021, Virtual Event, February 2-9, 2021,
  {AAAI} Press, 2021, pp. 6058--6065.

\bibitem{DBLP-icra21-learning}
M.~Iovino, J.~Styrud, P.~Falco, C.~Smith,
  \href{https://doi.org/10.1109/ICRA48506.2021.9562088}{Learning behavior trees
  with genetic programming in unpredictable environments}, in: {IEEE}
  International Conference on Robotics and Automation, {ICRA} 2021, Xi'an,
  China, May 30 - June 5, 2021, {IEEE}, 2021, pp. 4591--4597.
\newblock \href {https://doi.org/10.1109/ICRA48506.2021.9562088}
  {\path{doi:10.1109/ICRA48506.2021.9562088}}.
\newline\urlprefix\url{https://doi.org/10.1109/ICRA48506.2021.9562088}

\bibitem{DBLP-arxiv17-bt}
M.~Colledanchise, P.~{\"{O}}gren,
  \href{http://arxiv.org/abs/1709.00084}{Behavior trees in robotics and {AI:}
  an introduction}, CoRR abs/1709.00084 (2017).
\newblock \href {http://arxiv.org/abs/1709.00084} {\path{arXiv:1709.00084}}.
\newline\urlprefix\url{http://arxiv.org/abs/1709.00084}

\bibitem{Gupta_2023_CVPR}
T.~Gupta, A.~Kembhavi, Visual programming: Compositional visual reasoning
  without training, in: Proceedings of the IEEE/CVF Conference on Computer
  Vision and Pattern Recognition (CVPR), 2023, pp. 14953--14962.

\bibitem{WEB-control-groups}
{Linux Kernel}, \href{https://bit.ly/3A456Tp}{Control groups}, accessed:
  2023-06-18.
\newline\urlprefix\url{https://bit.ly/3A456Tp}

\bibitem{WEB-k8s-namespaces}
Kubernetes, \href{https://bit.ly/3A5skbE}{Namespaces}, accessed: 2023-06-18.
\newline\urlprefix\url{https://bit.ly/3A5skbE}

\bibitem{USENIX-atc22-rund}
Z.~Li, J.~Cheng, Q.~Chen, E.~Guan, Z.~Bian, Y.~Tao, B.~Zha, Q.~Wang, W.~Han,
  M.~Guo, Rund: {A} lightweight secure container runtime for high-density
  deployment and high-concurrency startup in serverless computing, in:
  J.~Schindler, N.~Zilberman (Eds.), 2022 {USENIX} Annual Technical Conference,
  {USENIX} {ATC} 2022, Carlsbad, CA, USA, July 11-13, 2022, {USENIX}
  Association, 2022, pp. 53--68.

\bibitem{WEB-otel}
OpenTelemetry, \href{https://opentelemetry.io}{High-quality, ubiquitous, and
  portable telemetry to enable effective observability}, accessed: 2023-06-18.
\newline\urlprefix\url{https://opentelemetry.io}

\bibitem{WEB-kubeedge}
KubeEdge, \href{https://kubeedge.io/}{A kubernetes native edge computing
  framework}, accessed: 2023-06-18.
\newline\urlprefix\url{https://kubeedge.io/}

\bibitem{WEB-k3s}
K3s, \href{https://k3s.io/}{Lightweight kubernetes}, accessed: 2023-06-18.
\newline\urlprefix\url{https://k3s.io/}

\bibitem{WEB-NATS-limit}
{NATS Docs}, \href{https://bit.ly/42WgGf9}{Configuration - limits}, accessed:
  2023-06-18.
\newline\urlprefix\url{https://bit.ly/42WgGf9}

\bibitem{WEB-Kafka-limit}
{Kafka Documentation}, \href{https://bit.ly/3XmBBH6}{Broker configs}, accessed:
  2023-06-18.
\newline\urlprefix\url{https://bit.ly/3XmBBH6}

\bibitem{WEB-nats}
NATS, \href{https://nats.io}{Connective technology for adaptive edge \&
  distributed systems}, accessed: 2023-06-18.
\newline\urlprefix\url{https://nats.io}

\bibitem{WEB-minio}
MinIO, \href{https://min.io}{Multi-cloud object storage}, accessed: 2023-06-18.
\newline\urlprefix\url{https://min.io}

\bibitem{WEB-prometheus}
Prometheus, \href{https://prometheus.io}{Monitoring system \& time series
  database}, accessed: 2023-06-18.
\newline\urlprefix\url{https://prometheus.io}

\bibitem{WEB-jaeger}
Jaeger, \href{https://www.jaegertracing.io}{open source, end-to-end distributed
  tracing}, accessed: 2023-06-18.
\newline\urlprefix\url{https://www.jaegertracing.io}

\bibitem{DBLP:journals/tpds/DengZXZJLYDZ22}
S.~Deng, H.~Zhao, Z.~Xiang, C.~Zhang, R.~Jiang, Y.~Li, J.~Yin, S.~Dustdar,
  A.~Y. Zomaya, \href{https://doi.org/10.1109/TPDS.2021.3137380}{Dependent
  function embedding for distributed serverless edge computing}, {IEEE} Trans.
  Parallel Distributed Syst. 33~(10) (2022) 2346--2357.
\newblock \href {https://doi.org/10.1109/TPDS.2021.3137380}
  {\path{doi:10.1109/TPDS.2021.3137380}}.
\newline\urlprefix\url{https://doi.org/10.1109/TPDS.2021.3137380}

\bibitem{DBLP:journals/percom/CicconettiCP22}
C.~Cicconetti, M.~Conti, A.~Passarella,
  \href{https://doi.org/10.1016/j.pmcj.2022.101689}{Faas execution models for
  edge applications}, Pervasive Mob. Comput. 86 (2022) 101689.
\newblock \href {https://doi.org/10.1016/j.pmcj.2022.101689}
  {\path{doi:10.1016/j.pmcj.2022.101689}}.
\newline\urlprefix\url{https://doi.org/10.1016/j.pmcj.2022.101689}

\bibitem{DBLP:journals/fgcs/PoojaraDJS22}
S.~R. Poojara, C.~K. Dehury, P.~Jakovits, S.~N. Srirama,
  \href{https://doi.org/10.1016/j.future.2021.12.012}{Serverless data pipeline
  approaches for iot data in fog and cloud computing}, Future Gener. Comput.
  Syst. 130 (2022) 91--105.
\newblock \href {https://doi.org/10.1016/j.future.2021.12.012}
  {\path{doi:10.1016/j.future.2021.12.012}}.
\newline\urlprefix\url{https://doi.org/10.1016/j.future.2021.12.012}

\bibitem{DBLP:journals/pieee/ZhouCLZLZ19}
Z.~Zhou, X.~Chen, E.~Li, L.~Zeng, K.~Luo, J.~Zhang,
  \href{https://doi.org/10.1109/JPROC.2019.2918951}{Edge intelligence: Paving
  the last mile of artificial intelligence with edge computing}, Proc. {IEEE}
  107~(8) (2019) 1738--1762.
\newblock \href {https://doi.org/10.1109/JPROC.2019.2918951}
  {\path{doi:10.1109/JPROC.2019.2918951}}.
\newline\urlprefix\url{https://doi.org/10.1109/JPROC.2019.2918951}

\bibitem{10128754}
X.~Deng, B.~Chen, X.~Chen, X.~Pei, S.~Wan, S.~K. Goudos, Trusted edge computing
  system based on intelligent risk detection for smart iot, IEEE Transactions
  on Industrial Informatics (2023) 1--10\href
  {https://doi.org/10.1109/TII.2023.3245681}
  {\path{doi:10.1109/TII.2023.3245681}}.

\bibitem{DBLP:conf/acsw/AslanpourTCJSTA21}
M.~S. Aslanpour, A.~N. Toosi, C.~Cicconetti, B.~Javadi, P.~Sbarski, D.~Taibi,
  M.~D. de~Assun{\c{c}}{\~{a}}o, S.~S. Gill, R.~Gaire, S.~Dustdar,
  \href{https://doi.org/10.1145/3437378.3444367}{Serverless edge computing:
  Vision and challenges}, in: N.~Stanger, B.~J. Woodford, M.~Winikoff, D.~M.
  Eyers, V.~L. Joachim, D.~A. da~Costa, A.~Trotman (Eds.), {ACSW} '21: 2021
  Australasian Computer Science Week Multiconference, Dunedin, New Zealand, 1-5
  February, 2021, {ACM}, 2021, pp. 10:1--10:10.
\newblock \href {https://doi.org/10.1145/3437378.3444367}
  {\path{doi:10.1145/3437378.3444367}}.
\newline\urlprefix\url{https://doi.org/10.1145/3437378.3444367}

\bibitem{9931467}
R.~Xie, D.~Gu, Q.~Tang, T.~Huang, F.~R. Yu, Workflow scheduling in serverless
  edge computing for the industrial internet of things: A learning approach,
  IEEE Transactions on Industrial Informatics 19~(7) (2023) 8242--8252.
\newblock \href {https://doi.org/10.1109/TII.2022.3217477}
  {\path{doi:10.1109/TII.2022.3217477}}.

\bibitem{DBLP:conf/ccgrid/GackstatterFD22}
P.~Gackstatter, P.~A. Frangoudis, S.~Dustdar,
  \href{https://doi.org/10.1109/CCGrid54584.2022.00023}{Pushing serverless to
  the edge with webassembly runtimes}, in: 22nd {IEEE} International Symposium
  on Cluster, Cloud and Internet Computing, CCGrid 2022, Taormina, Italy, May
  16-19, 2022, {IEEE}, 2022, pp. 140--149.
\newblock \href {https://doi.org/10.1109/CCGrid54584.2022.00023}
  {\path{doi:10.1109/CCGrid54584.2022.00023}}.
\newline\urlprefix\url{https://doi.org/10.1109/CCGrid54584.2022.00023}

\bibitem{ZHENG2023105}
S.~Zheng, B.~Liu, W.~Lin, X.~Ye, K.~Li, A package-aware scheduling strategy for
  edge serverless functions based on multi-stage optimization, Future
  Generation Computer Systems 144 (2023) 105--116.
\newblock \href {https://doi.org/https://doi.org/10.1016/j.future.2023.02.013}
  {\path{doi:https://doi.org/10.1016/j.future.2023.02.013}}.

\end{thebibliography}
\appendix
\section{A Real-world Use Case of BeeFlow}
\label{appendix:real-world}

In this section, we present a real-world use case of \emph{BeeFlow} to help readers better
capture the composite features of behavior trees and gain a clear idea of how
\emph{BeeFlow} can be beneficial.
This example focuses on large language model (LLM)-based domain-specific program generation,
which is an on-going project within our team.
The reasons we have chosen this example are two-fold:
(1) LLMs are increasingly garnering attention and becoming more useful, and many
readers may already be familiar with related concepts;
(2) this example has been developed for internal use in our evaluation of different LLMs for
domain-specific program generation and has proven useful for generating
feasible programs and collecting datasets for model fine-tuning.

\begin{figure*}[!ht]
    \centering
    \includegraphics[width=\linewidth]{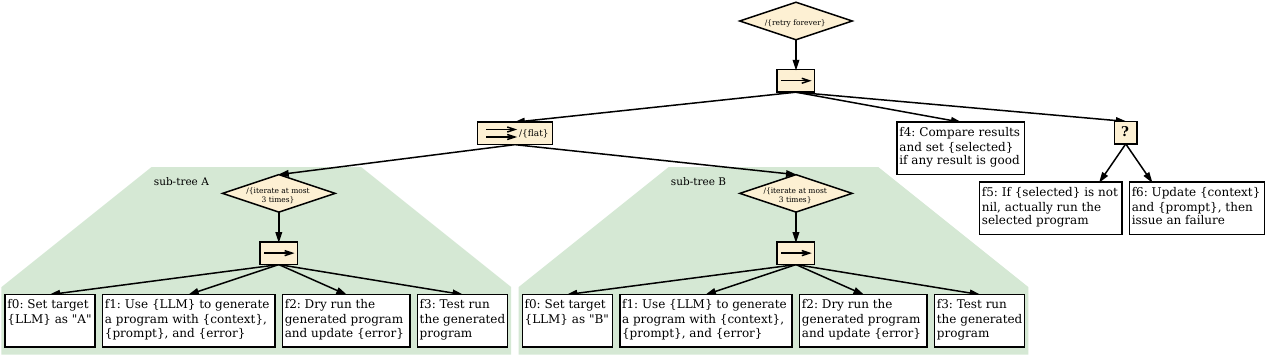}
    \vspace{-0.5em}
    \caption{A real-world use case of \emph{BeeFlow} for LLM-based
             domain-specific program generation.}
    \label{fig:real-world}
\end{figure*}

Essentially, this example follows the idea proposed in \textsc{VisProg} \cite{Gupta_2023_CVPR}
for program generation,
where a target instruction (prompt) and in-context program generation samples (context)
are provided to an LLM to generate programs that meet the requirements of the instruction.
Although LLMs excel in natural language understanding and content generation, they can
occasionally output invalid programs (e.g., using a non-existent command in the generated
program). To address this issue, we need an external workflow to eliminate invalid programs
and enhance program quality. We have constructed the workflow using \emph{BeeFlow}, and
a simplified version for demonstration purposes is displayed in Figure~\ref{fig:real-world}.

In subtree A, depicted in Figure~\ref{fig:real-world}, following the program generation by the LLM,
a dry run is scheduled to ensure that basic constraints for the domain-specific program
are satisfied. For instance, the generated program must not use non-existent commands, and its
syntax must adhere to the definition of the corresponding domain-specific language. If the dry run
fails, an error message will be generated, and the generation process will repeat, provided with
the error message that enables the LLM to understand its mistakes and potentially avoid them
in the next generation. A test run is issued only if the generated program passes the dry run test.
Moreover, a decorator composite controls the maximum number of iterations/retries for program
generation.

Subtree A and subtree B in Figure~\ref{fig:real-world} are fundamentally the same, with
the exception that they utilize different LLMs. Running multiple LLMs is necessary because we
aim to fine-tune a smaller LLM suitable for resource-constrained edge devices by learning from
a larger LLM. Being able to run multiple LLMs concurrently allows us to collect samples for
fine-tuning efficiently.

Once all subtrees complete their tasks (either generating a feasible program or exhausting
the iteration limit), an evaluation process scores the results obtained from feasible programs.
A program is selected if its result meets the requirements of the instruction effectively.

Finally, if a program is selected, an actual run can be scheduled. Generally, an actual run is
more resource-intensive than a test run. For instance, in an image editing scenario, an actual run
may require the highest image resolution for generation, whereas a test run might suffice a lower
resolution. In a robotics scenario, an actual run may involve real robots, while a test run could
simulate robots in a virtual environment. If no program is selected, i.e., none of the results
from the generated programs fulfill the instruction, human intervention is typically
needed to update the context and prompt for LLMs. The whole process can repeat until
a satisfactory result is achieved. All functions involved in the process are listed in
Table~\ref{tab:real-world-functions} for easier retrieval.

\begin{table}[!ht]
    \centering
    \caption{Functions used in LLM-based domain-specific program generation.}
    \label{tab:real-world-functions}
    \vspace{-0.5em}\small
    \begin{tabular}{|p{0.8cm}|p{4.2cm}|p{2.4cm}|}
        \hline
        \textbf{Name}
        & \textbf{Description}
        & \textbf{Failure Condition}
        \\\hline\hline
        f0
        & Set the target LLM for program generation
        & Unexpected
        \\\hline
        f1
        & Use the target LLM to generate a domain-specific program
            given the context, prompt, and an optional error message
        & Unexpected
        \\\hline
        f2
        & Dry run the generated program to detect errors like the use of
            non-existent commands, wrong syntax, etc
        & When dry run fails
        \\\hline
        f3
        & Test run the generated program with cheap operations
        & When test run fails
        \\\hline
        f4
        & Use domain-specific tools to compare results obtained from the
            generated programs. If there are results that are considered good
            enough, select the program corresponding to the best result
        & Unexpected
        \\\hline
        f5
        & Run the selected program with standard operations
        & When selected is empty or run fails
        \\\hline
        f6
        & Update context and/or prompt, usually requires human efforts
        & Always
        \\\hline
        \{flat\}
        & Insert resulted \emph{subtree} payloads directly to the main payload
        & Unexpected
        \\\hline
    \end{tabular}
\end{table}

The behavior tree exhibited in Figure~\ref{fig:real-world} orchestrates all the procedures
we described above in a clean and organized manner. Based on our experience using \emph{BeeFlow}
in our LLM-based domain-specific program generation project, we have identified the following
pros and cons:

\textbf{Pros}\begin{itemize}
    \item A new comer to the project can quickly grasp the overall idea
        without delving into project details.
    \item While a developer works on a \emph{subtree} or a single \emph{leaf}, it understands
        how the \emph{subtree} or \emph{leaf} interacts with other parts, again, without diving
        into the details of those parts. This understanding has fostered better exchanges
        within our project.
    \item Versioning \emph{subtrees} and \emph{leaves} has become simpler, given that behavior trees
        are modular and hierarchical at all scales.
    \item Resource mapping can be optimized automatically.
\end{itemize}

\textbf{Cons}\begin{itemize}
    \item Although behavior trees are generally easy to comprehend, it can be challenging
        for a beginner to consider and compose things in a behavior tree manner.
    \item Payload transmitted among nodes in behavior trees should generally be made explicit,
        which can be cumbersome in some cases.
\end{itemize}

Despite the additional efforts required for the adoption of \emph{BeeFlow},
we believe that the modularity, hierarchy, and automatic mapping features it offers will ultimately
make continuously growing projects easier to maintain.

\section{Evaluation of BeeFlow in a Conventional Edge Environment}
\label{appendix:edge-con}

To better understand the impact of different parameter configurations of testbeds on the performance
of \emph{BeeFlow}, we present an additional evaluation in a conventional edge environment
in this section.

Compared with a high-density, resource-constrained edge testbed, a conventional edge testbed
will typically features powerful processors, larger memory capacity, and faster storage and
networking devices. These improved parameters suggest that a conventional edge testbed can
achieve shorter processing latency and higher processing throughput. However, these enhancements
come at the cost of both capitals and operations, leading to a smaller number of
nodes in a conventional edge environment in general.

Moreover, when compared with a high-profile cloud testbed, a conventional edge testbed usually
has lower-profile processors, which means fewer number of cores, smaller caches and memory
bandwidth, and a reduced number of PCIe lanes. Nevertheless, unlike cloud servers that are
designed for fine-grained virtualization, both computing processors and peripheral devices,
such as storage and networking devices, are dedicated in conventional edge servers. These
differences suggest that server nodes in a conventional edge environment are less
scalable than cloud servers but may deliver comparable node-wise performance.

Table~\ref{tab:conventional} displays the detailed specifications of the conventional edge
environment we utilized, consisting of 4 Dell T5820 workstations connected through a 10GbE switch.

\begin{table}[!ht]
    \centering
    \caption{Specifications of nodes in the conventional edge testbed.}
    \vspace{-0.5em}
    \label{tab:conventional}
    \begin{tabular}{@{}R{2.5cm}|L{5.5cm}@{}}
        \toprule
        \textbf{1*storage node \emph{\footnotesize (\tFaaSFlow only)}}
        & Intel Xeon W-2145,
        \mbox{8 cores / 16 threads}, \mbox{32GB RAM}, 250GB NVMe SSD
        \\\midrule
        \textbf{3*worker node}
        & Intel Xeon W-2145,
        \mbox{8 cores / 16 threads}, \mbox{32GB RAM}, 250GB NVMe SSD
        \\\bottomrule
    \end{tabular}
\end{table}

\begin{figure}[!ht]
    \centering
    \includegraphics[scale=0.75]{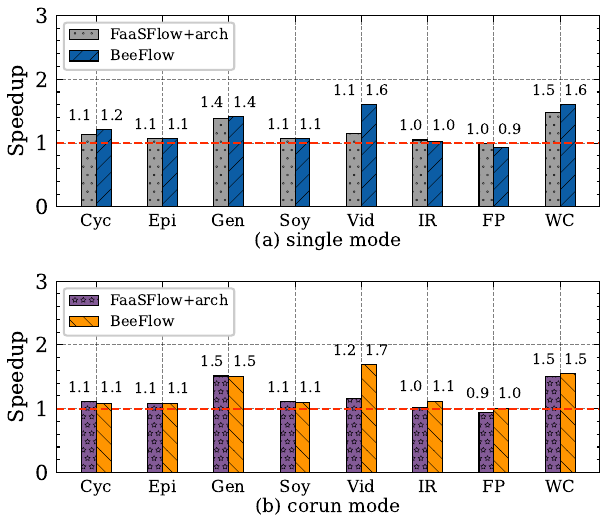}
    \vspace{-0.5em}
    \caption{
        Speedup over \emph{FaaSFlow} in both modes, respectively (conventional edge).
    }
    \label{fig:e2e-con-edge}
\end{figure}

We follow the same deployment setup as the cloud testbed for the conventional edge testbed, and
the speedup over \emph{FaaSFlow} achieved by \emph{BeeFlow} is illustrated in
Figure~\ref{fig:e2e-con-edge}.

A notable difference between the results obtained from the conventional edge testbed and
the cloud testbed is that \emph{FaaSFlow+arch} does not degenerate in the Cyc workflow and even
achieves a small speedup. This observation can be attributed to the fact that the conventional
edge testbed has direct, dedicated storage and networking devices, in contrast to the cloud testbed
where storage and networking devices are virtualized for sharing across many cloud instances.
Dedicated storage and networking devices generally deliver more deterministic I/O performance
and lower overhead, compensating for the absence of in-memory caching in \emph{FaaSFlow+arch}.

Although \emph{FaaSFlow+arch} and \emph{BeeFlow} exhibit better or comparable performance
over \emph{FaaSFlow} in all benchmark workflows in the conventional edge testbed,
the improvement of \emph{BeeFlow} over \emph{FaaSFlow+arch} is not as pronounced as in the
high-density edge testbed and high-profile cloud testbed. This is because the conventional edge
testbed has only 3 worker nodes, offering limited opportunities for \emph{BeeFlow} to optimize
from the resource contention aspect. Nonetheless, \emph{BeeFlow} still achieves performance
comparable with the placement scheme resulting from the critical path-based method employed by
\emph{FaaSFlow}, which confirms the robustness of \emph{BeeFlow} in various environments.

In conclusion, the additional evaluation in a conventional edge environment suggests that the
adoption of \emph{BeeFlow} will yield greater benefits when there is ample room for mitigating
resource contention (i.e., when the scale of the environment is sufficiently large).
However, even in small-scale environments, \emph{BeeFlow} can still achieve moderate
performance, demonstrating its good robustness.

\end{document}